\documentclass[aps,prd,showpacs,floatfix,preprintnumbers,nofootinbib,amssymb,amsmath]{revtex4}
\usepackage{graphicx,bm,color}
\def \l {\left}
\def \r {\right}
\def \la {\left\langle}
\def \ra {\right\rangle}
\def \bit{\begin{itemize}}
\def \eit{\end{itemize}}
\def \ben{\begin{enumerate}}
\def \een{\end{enumerate}}
\def \bsubeq{\begin{subequations}}
\def \esubeq{\end{subequations}}
\def \beq{\begin{equation}}
\def \eeq{\end{equation}}
\def \beqa{\begin{eqnarray}}
\def \eeqa{\end{eqnarray}}
\newcommand{\dslash}{\partial\!\!\!/}
\newcommand{\Dslash}{D\!\!\!\!/}

\def \etal{{\sl et al.~\/}}

\def \eup{{\sl Eur.\ Phys. ~\/}}
\def \jhep{{\sl J.\ H.\ E.\ P.~\/}}

\def \jphys{{\sl J.\ Phys.~\/}}

\def \nc{{\sl Nuovo\ Cim.\ ~\/}}
\def \np{{\sl Nucl.\ Phys.~\/}}

\def \pl{{\sl Phys.\ Lett.~\/}}
\def \pr{{\sl Phys.\ Rev.~\/}}
\def \prp{{\sl Phys.\ Rept.~\/}}
\def \prgnp{{\sl Prog.\ Part.\ Nucl.\ Phys.~\/}}
\def \prl{{\sl Phys.\ Rev.\ Lett.~\/}}
\def \rmp{{\sl Rev.\ Mod.\ Phys.~\/}}
\def \eg{{\sl e.g.\/}}
\def \ie{{\sl i.e.\/}}
\def \eg{{\sl e.g.\/}}
\def \etal{{\sl et al.\/}}

\begin{document}
 
\title{Thermodynamics of the PNJL model with nonzero baryon 
and isospin chemical potentials}
\author{Swagato \surname{Mukherjee}}
\email{swagato@tifr.res.in}
\affiliation{Department of Theoretical Physics, \\ Tata Institute,
of Fundamental Research, Homi Bhabha Road, Mumbai 400 005, India.}
\author{Munshi G. \surname{Mustafa}}
\email{munshigolam.mustafa@saha.ac.in}
\author{Rajarshi \surname{Ray}}
\email{rajarshi.ray@saha.ac.in}
\affiliation{Theory Division, Saha Institute of Nuclear Physics, %\\
          1/AF, Bidhannagar, Kolkata 700 064, India.}

\begin{abstract}

We have extended the Polyakov-Nambu-Jona-Lasinio (PNJL) model  for two
degenerate flavours to include the isospin chemical potential ($\mu_I$). 
All the diagonal and mixed derivatives of pressure with respect to the 
quark number (proportional to baryon number) chemical potential ($\mu_0$)
and isospin chemical potential upto sixth order have been extracted at 
$\mu_0 = \mu_I = 0$. These derivatives give the generalized 
susceptibilities with respect to quark and isospin numbers. Similar 
estimates for the flavour diagonal and off-diagonal susceptibilities 
are also presented. Comparison to Lattice QCD (LQCD) data of some of 
these susceptibilities for which LQCD data are available, show similar 
temperature dependence, though there are some quantitative
deviations above the crossover temperature. We have also looked at the 
effects of instanton induced flavour-mixing coming from the $U_A(1)$ 
chiral symmetry breaking 't Hooft determinant like term in the NJL part 
of the model. The diagonal quark number and isospin susceptibilities
are completely unaffected. The off-diagonal susceptibilities show
significant dependence near the crossover. 
Finally we present the chemical potential dependence of specific heat 
and speed of sound within the limits of chemical potentials where neither 
diquarks nor pions can condense.

\end{abstract}
\preprint{TIFR/TH/06-28, SINP/TNP/06-25}
\maketitle

\section{Introduction} \label{sc.intro}

Two most important features of strongly interacting matter at low
temperature and chemical potentials are the phenomenon of color charge
confinement and chiral symmetry breaking. However, with the increase in
temperature and/or chemical potential, various phases may appear with
different confining and chiral properties. At present both theoretical
and experimental endeavours are underway to map out the phase diagram of
QCD.

In the limit of infinite quark mass, the thermal average of the
Polyakov-loop can be considered as the order parameter for the
confinement-deconfinement transition \cite{polyl}. Though in presence of
dynamical quarks the Polyakov-loop is not a rigorous order parameter for
this transition, it still serves as an indicator of a rapid quark-hadron
crossover. Motivated by this observation, Polyakov-loop based effective
theories have been suggested \cite{polyd1,polyd2,polyd3} to capture the
underlying physics of the confinement-deconfinement transition. The
essential ingredient of these models is an effective potential
constructed out of the Polyakov-loop (and its complex conjugate). More
recently, the parameters in these effective theories have been fixed
\cite{pisarski1,pisarski2} using the data from Lattice QCD (LQCD)
simulations (similar comparisons of perturbative effects on
Polyakov-loop with Lattice data above the deconfinement transition
was studied in \cite{salcd}).

With the small quark masses the QCD Lagrangian has a partial global
chiral symmetry, which is however broken spontaneously at low
temperatures (and hence the absence of chiral partners of low-lying
hadrons). This symmetry is supposed to be partially restored at 
higher temperatures and chemical potentials. The chiral condensate is 
considered to be the order parameter in this case. Various effective 
chiral models exist for the study of physics related to the chiral 
dynamics, e.g. the sigma model \cite{sgmodel} and the Nambu-Jona-Lasinio 
(NJL) model \cite{njl1,njls}. The parameters of these models are 
fixed from the phenomenology of the hadronic sector.

Various studies of the QCD inspired models
indicate (see e.g. Refs.\cite{qcdpd1,qcdpd2,qcdpd3,qcdpd4,qcdpd5})
that at low temperatures there is a possibility of first order phase
transition for a large baryon chemical potential $\mu_{B_c}$. This
$\mu_{B_c}$ is supposed to decrease with increasing temperature. Thus
there is a first order phase transition line starting from
($T=0$, $\mu_B=\mu_{B_c}$) on the $\mu_B$ axis in the ($T$,$\mu_B$)
phase diagram which steadily bends towards the ($T=T_c$, $\mu_B=0$)
point and may actually terminate at a critical end point (CEP)
characterized by ($T=T_E$, $\mu_B=\mu_{B_E}$), which can be detected via
enhanced critical fluctuations in heavy-ion reactions \cite{rajag1}.
The location of this CEP has become a topic of major importance in
effective model studies (see e.g. Ref.\cite{njltype}). For $\mu_B \ne 0$
LQCD has a complex determinant which hinders usual importance sampling
techniques. However recently the CEP was located for the physical 
\cite{fdkz1} and for somewhat larger \cite{fdkz2} quark masses using 
the reweighting technique of \cite{smalmu1}, and for Taylor expansion 
method in \cite{ceprs}.

For nonzero isospin chemical potential ($\mu_I$) models and effective 
theories \cite{muimod} find an interesting array of possible phases.
The most important phenomenon that is supposed to happen is a 
transition to the pion condensed phase close to $\mu_I \sim m_{\pi}$. This
has also been supported by Lattice simulations \cite{muilat}, which 
does not suffer from the complex determinant problem for $\mu_I \ne 0$ 
and $\mu_B = 0$. 

In this paper we study some of the thermodynamic properties of strongly
interacting matter using the Polyakov loop enhanced Nambu-Jona-Lasinio 
(PNJL) model \cite{pnjl0,pnjl1}. In this model one is able to couple 
the chiral and deconfinement order parameters inside a single framework.  
While the NJL part is supposed to give the correct chiral properties,
the Polyakov-loop part simulates the deconfinement physics.
In fact studies of Polyakov loop coupled to chiral quark models have
become quite fashionable these days (see e.g. Ref.\cite{ruiz1}).

The
initial motivation to couple Polyakov loop to the NJL model was to
understand the coincidence of chiral symmetry restoration and
deconfinement transitions observed in LQCD simulations \cite{coinlat}.
While the NJL part is supposed to give the correct chiral properties,
the Polyakov-loop part simulates the deconfinement physics.  Indeed the
PNJL model worked well to obtain the ``coincidence" of onset of chiral
restoration and deconfinement \cite{pnjl0,pnjl1}.  Recently the
introduction of the Polyakov loop potential \cite{pnjl1,pnjl2} has made
it possible to extract estimates of various thermodynamic quantities.
The pressure, scaled pressure difference at various quark chemical
potential $\mu_0$ (or baryon chemical potential $\mu_B$, where 
$\mu_B=3 \mu_0$), quark number density and the interaction measure were
extracted from the PNJL model in Ref. \cite{pnjl2} for two quark
flavours, and all the quantities compared well with the LQCD data.
Following this some of us made a comparative study \cite{pnjl3} of the
quark number susceptibility (QNS) and its higher order derivatives with
respect to $\mu_0$ with LQCD data. Here the qualitative features match
very well though there are some quantitative differences. Very recently 
the spectral properties of low lying meson states have been studied 
in \cite{ratti2}.

Encouraged by these results, in this paper we have
extended the the PNJL model to incorporate the effects of nonzero
isospin chemical potential ($\mu_I$). The motivation for this is that,
it enables one to calculate the isospin number susceptibility (INS)
and its higher order derivatives with respect to $\mu_0$. LQCD data on
these quantities are also available \cite{sixx}. Thus comparing the
results of PNJL for these quantities with that for the LQCD data will
provide an opportunity to perform some stringent tests on the PNJL model.  

Moreover, once both the QNS and the INS are known one can proceed
further to compute the flavour diagonal and off-diagonal
susceptibilities separately. Since the $2$-nd order flavour off-diagonal
susceptibility measures the correlation among ``up'' ($u$) and ``down''
($d$) flavours \cite{gavai}, this quantity provide a direct
understanding to the extent in which the PNJL model captures the
underlying physics of QCD.

In our attempt to have a closer look at the $u$-$d$ flavour correlation
within the PNJL model, we have modified the NJL part of the
PNJL model by using the NJL Lagrangian proposed in
\cite{buballa}. This Lagrangian has a term that can be interpreted as an
interaction induced by instantons and reflects the $U_A(1)$-anomaly of
QCD. It has the structure of a 't-Hooft determinant in the
flavour space, leading to flavour-mixing. By adjusting the relative
strength of this term one can explicitly control the amount of
flavour-mixing in the NJL sector. This modified NJL Lagrangian reduces
to the standard NJL Lagrangian \cite{njl1,njls} in some particular
limit. This modification of the PNJL model have allowed us to study the
effects of such flavour-mixing on various susceptibilities, specially on
the $2$-nd order off-diagonal one which measures the $u$-$d$ flavour
correlation.

Investigation of the flavour-mixing effects brings us to an
important issue regarding the NJL-type models. Within
the framework of an NJL model it has been found \cite{asakawa1} that for
$\mu_I=0$, in the $T-\mu_0$ plane, there is a single first order phase
transition line (which ends at a critical endpoint) at low temperatures.
But for $\mu_I\ne0$ this single line separates into two first order
phase transition lines because of the different behaviour of the $u$ and
$d$ quark condensates \cite{toubl1}. Thus there is a possibility of
having two critical end-points in the QCD phase diagram \cite{toubl1}.
This has also been observed in Random Matrix models \cite{verba1}, in
ladder QCD models \cite{laddq1} as well as in hadron resonance gas
models \cite{hadres1}. It was then argued in Ref. \cite{buballa,zhuang2} that 
the flavour-mixing through the instanton effects \cite{rein1,bernard1}
may wipe out this splitting. Later studies found that the splitting is
considerable when $\mu_I$ is large \cite{hadres1,barducci1} or $\mu_B$
is large \cite{toubl2}. We shall restrict ourselves only to small 
chemical potentials and calculate the susceptibilities with the 
modified PNJL model for different amount of flavour-mixing. Comparing 
these with LQCD data may give us some idea about actual amount of the 
flavour-mixing that is favoured by the LQCD simulations.

Our next objective is to study the specific heat at constant volume 
($C_V$) and speed of sound ($v_s$) of strongly interacting systems. 
These two quantities are of major importance for heavy-ion collision
experiments. While $C_V$ is related to the event-by-event temperature
fluctuations \cite{ebe1} and mean transverse momentum fluctuations
\cite{ebe2} in heavy-ion collisions, the quantity $v_s$ controls the
expansion rate of the fireball produced in such collisions and hence an
important input parameter for the hydrodynamic studies
\cite{elp1,elp2,elp3,elp4}. The temperature dependence of these
quantities were reported earlier in Ref.\ \cite{pnjl3}. For the sake of
completeness, in this paper we have also studied the quark number and
isospin chemical potential dependence of $C_V$ and $v_s$.

The plan of this paper is as follows. In Section \ref{sc.formal}, we
will present our formalisms. First, we will briefly discuss   the
extended PNJL model which we are going to use. Next, in the same
section, formalisms regarding the Taylor expansion of pressure (with
respect to $\mu_0$ and $\mu_I$) and formulae for specific heat $C_V$ and
speed of sound $v_s$ will be given. In Section\ \ref{sc.results} we will
present our results and compare some of those with the available LQCD
data.  Finally, we conclude with a discussion in Section\
\ref{sc.summary}.  Detail mathematical expressions regarding the model
can be found in Appendix\ \ref{ap.model}. 

\section{Formalism} \label{sc.formal}
\subsection {PNJL Model}
\label{sc.pnjl}

The PNJL model at nonzero temperature $T$ and quark number chemical
potential $\mu_0$ was introduced in Ref.\ \cite{pnjl1,pnjl2}. Here we 
extend it to include the isospin chemical potential $\mu_I$. We have 
introduced separate chemical potentials 
$\mu_u$ and $\mu_d$ for the ``up" and ``down" quark flavours respectively
in the NJL model following Ref.\ \cite{asakawa1,buballa}. 
To further extend it to include the Polyakov loop dynamics we have 
followed the parameterization of the PNJL model used in 
Ref.\ \cite{pnjl2}. We start with the final form of the mean field 
thermodynamic potential per unit volume that we have obtained.
It is given by (further details about the model can be found in
Appendix\ \ref{ap.model}), 
\beqa
\Omega&=&{\cal U}\left(\Phi,\bar{\Phi},T\right)+
2 G_1(\sigma_u^2 + \sigma_d^2) + 4 G_2 \sigma_u \sigma_d 
\nonumber \\
&-& \sum_{f=u,d}
2\,T\int\frac{\mathrm{d}^3p}{\left(2\pi\right)^3}
\left\{ \ln\left[1+3\left(\Phi+\bar{\Phi}\mathrm{e}^
{-\left(E_f-\mu_f\right)/T}\right)\mathrm{e}^{-\left(E_f-\mu_f\right)/T}
 + \mathrm{e}^{-3\left(E_f-\mu_f\right)/T}\right]\right . \nonumber\\
&+& 
\left . 
\ln\left[1+3\left(\bar{\Phi}+\Phi\mathrm{e}^{-\left(E_f+\mu_f\right)/T}
\right)\mathrm{e}^{-\left(E_f+\mu_f\right)/T}+
\mathrm{e}^{-3\left(E_f+\mu_f\right)/T}\right] \right\}
- \sum_{f=u,d} 6\int\frac{\mathrm{d}^3p}{\left(2\pi\right)^3}{E_f}
\theta\left(\Lambda^2-\vec{p}^{~2}\right) ~~~.
\label{omega}
\eeqa
Here for the two flavours the respective quark condensates are given by 
$\sigma_u = <\bar{u}u>$ and $\sigma_d=<\bar{d}d>$ 
%%%%%%%%%%%%%%%%%%%%%%%%%%%%%%%%%%%%%%%%%%%%%%%%%%%%%%%%%%%%%%%%%%%%%%%
\footnote{Here we deviate from the convention of defining the sigma
condensates from those of Refs. \cite{pnjl2,pnjl3}.}
%%%%%%%%%%%%%%%%%%%%%%%%%%%%%%%%%%%%%%%%%%%%%%%%%%%%%%%%%%%%%%%%%%%%%%%
and the respective
chemical potentials are $\mu_u$ and $\mu_d$. Note that 
$\mu_0 = (\mu_u+\mu_d)/2$ and $\mu_I = (\mu_u-\mu_d)/2$. The quasi-particle 
energies are $E_{u,d}=\sqrt{\vec{p}^{~2}+m_{u,d}^2}$, where 
$m_{u,d}=m_0-4 G_1 \sigma_{u,d} -4 G_2 \sigma_{d,u}$ are the 
constituent quark masses and $m_0$ is the current quark mass
(we assume flavour degeneracy).  $G_1$ and $G_2$ are the effective 
coupling strengths of a local, chiral symmetric four-point 
interaction. $\Lambda$ is the 3-momentum cutoff in the NJL model. 
${\cal U}\left(\Phi,\bar{\Phi},T\right)$ is the effective
potential for the mean values of the traced Polyakov-loop $\Phi$ and 
its conjugate $\bar{\Phi}$, and $T$ is the temperature. The functional 
form of the potential is,

\beqa
\frac{\mathcal{U}\left(\Phi,\bar{\Phi},T\right)}{ T^4} =
-\frac{b_2\left(T\right)}{ 2 }\bar{\Phi} \Phi-
{b_3\over 6}\left(\Phi^3+
{\bar{\Phi}}^3\right)+ {b_4\over 4}\left(\bar{\Phi} \Phi\right)^2 ~~~,
\label{uu}
\eeqa

\noindent
with

\beqa
b_2\left(T\right)=a_0+a_1\left(\frac{T_0}{T}\right)
+a_2\left(\frac{T_0}{T} \right)^2+a_3\left(\frac{T_0}{T}\right)^3~~~.
\label{bb}
\eeqa

\noindent
The coefficients $a_i$ and $b_i$ were fitted from LQCD data of
pure gauge theory. The parameter $T_0$ is precisely the transition
temperature for this theory, and as indicated by LQCD data its
value was chosen to be $270\,{\rm MeV}$ \cite{tcpg1,tcpg2,tcpg3}. With the 
coupling to NJL model the transition doesn't remain first order. 
In this case from the peak in $d\Phi/dT$ the transition (or crossover) 
temperature $T_c$ comes around $227\,{\rm MeV}$. 

Before we move further we note some important features of this model:

\begin{itemize}

\item
Since the gluons in this model are contained only in a static 
background field, the model would be suitable to study the physics 
below $T = 2.5 T_c$. Above this temperature the
transverse degrees of freedom become important \cite{tranv}. 

\item 
In general, pion condensation takes place in NJL models
for $\mu_I > m_{\pi}/2$. Also there is a chiral transition
for $\mu_0 \sim 340\, {\rm MeV}$ above which diquark physics
become important. For simplicity we neglect both
the pion condensation and diquarks 
%%%%%%%%%%%%%%%%%%%%%%%%%%%%%%%%%%%%%%%%%%%%%%%%%%%%%%%%%%%%%%%%%
\footnote{Very recently diquarks have been discussed in Ref.
\cite{ratti4} and pion condensation in Ref.\cite{zliu}.}
%%%%%%%%%%%%%%%%%%%%%%%%%%%%%%%%%%%%%%%%%%%%%%%%%%%%%%%%%%%%%%%%%
and so restrict our
analysis to $\mu_I < 70\, {\rm MeV}$ and $\mu_0 < 200\, {\rm MeV}$.

\item

As discussed in the appendix, for $G_2=0$ the full symmetry of the 
Lagrangian in the chiral
limit ($m_0 = 0$) is $SU_V(2) \times SU_A(2) \times U_V(1) 
\times U_A(1)$. The coefficient $G_2$ is interpreted as inducing 
instanton effects as it breaks the $U_A(1)$ symmetry explicitly
by mixing the quark flavours.  By using a parameterization 
$G_1 = (1-\alpha)G_0$ and $G_2 = \alpha G_0$ (following
Ref.\  \cite{buballa}), one can tune the amount of instanton induced 
flavour mixing by varying $\alpha$. For $\alpha=0$ there 
is no instanton induced flavour mixing, and for $\alpha=1$ the
mixing becomes maximal. We shall look into the effects of this
mixing in the susceptibilities.

The form of the NJL part in Eqn.\ (\ref{omega}) is a generalization of 
the standard NJL model, which we get when $G_1 = G_2$, and 
$\mu_u = \mu_d$. In fact the potential in Eqn.\ (\ref{omega}) becomes exactly
the same as that of Ref.\ \cite{pnjl2,pnjl3} if we use $\alpha=0.5$
and put $G_0$ equal to half the four-point coupling $G$ in those
references. We shall use this value for $G_0$ in this work. 

\item
For the NJL sector without coupling to the Polyakov loop 
(i.e. setting $\Phi = \bar{\Phi} = 1$) one can easily see that
the expression for $\Omega$ in Eqn.\ (\ref{omega}), is invariant under 
the transformations $\mu_u \rightarrow -\mu_u$ ``and/or"
$\mu_d \rightarrow -\mu_d$. This implies that the physics along
the directions of $\mu_0 = 0$ and $\mu_I = 0$ at any given temperature
are equivalent. However inclusion of the Polyakov loop turns
off this symmetry. Now $\Omega$ is invariant only under the 
simultaneous transformation $\mu_u \rightarrow -\mu_u$ ``and"
$\mu_d \rightarrow -\mu_d$. This is a manifestation of the
CP symmetry which implies that $\Omega$ is symmetric only under the 
simultaneous transformation $\Phi \rightarrow \bar{\Phi}$
and $\mu_{u,d} \rightarrow -\mu_{u,d}$ and vice-verse. Thus 
coefficients of $\Phi$ and $\bar{\Phi}$ are found to be equal when
$\mu_0=0$, and different when $\mu_I=0$. In the $T - \mu_I$ plane we 
shall have $\Phi = \bar{\Phi}$, and everywhere else $\Phi \ne \bar{\Phi}$.
This is reminiscent of the complex fermion determinant for nonzero
$\mu_0$. This will be seen to have important consequences for the 
extraction of susceptibilities.

\item
On the other hand the quark condensates $\sigma_u$ and $\sigma_d$ are
equal to each other whenever $\mu_0 = 0$ or $\mu_I = 0$ [see, \eg, Eqn.\
(\ref{eq.sig-u-d})]in the NJL as well as PNJL model. This can be seen by
inspecting the thermodynamic potential $\Omega$ and remembering that we
are using $G_1 + G_2 = G_0$, and also the fact that for $\mu_0 = 0$,
$\Phi = \bar{\Phi}$.  Now $G_1$ and $G_2$ are only coupled to the
$\sigma_u$ and $\sigma_d$.  It is clear from Eqn.\ (\ref{omega}) that
whenever $\sigma_u = \sigma_d$, the couplings $G_1$ and $G_2$ come in 
the combination
$G_1 + G_2 = G_0 = {\rm constant}$.  This means that the physics is
completely independent of these couplings whenever either $\mu_0 = 0$
or $\mu_I = 0$.

\end{itemize}

\subsection{Taylor expansion of Pressure}
\label{sc.tayexp}

The pressure as a function of temperature $T$, quark chemical 
potential $\mu_0$ and isospin chemical potential $\mu_I$ is given by,

\beqa
P(T,\mu_0,\mu_I) = -\Omega(T,\mu_0,\mu_I) ~~~.
\label{prsu}
\eeqa

\noindent
Following usual thermodynamical relations one can show that the first 
derivative of pressure with respect to $\mu_0$ gives the quark
number density. The second derivative is the quark number susceptibility.
In LQCD since usual Monte Carlo importance sampling
fails for nonzero $\mu_0$, the QNS and higher order derivatives 
computed at $\mu_0=0$ can be used as Taylor expansion coefficients to 
extract chemical potential dependence of pressure. 

Given the thermodynamic potential $\Omega$, our job is to minimize it with 
respect to the fields $\sigma_u$, $\sigma_d$, $\Phi$ and $\bar{\Phi}$, 
using the following set of equations,

\beqa
{\partial\Omega\over\partial\sigma_u} = 0 ~,~~~~~
{\partial\Omega\over\partial\sigma_d} = 0 ~,~~~~~
{\partial\Omega\over\partial\Phi} = 0 ~,~~~~~
{\partial\Omega\over\partial \bar{\Phi}} = 0 ~.
\label{solv}
\eeqa

The values of the fields so obtained can then be used to evaluate all 
the thermodynamic quantities in mean-field approximation. The cross-over
temperature for $\mu_0=\mu_I=0$ was obtained in Ref.\  \cite{pnjl3} and 
was found to be $T_c = 227\, {\rm MeV}$.
The field values obtained from Eqn.\ (\ref{solv}) are then put back into 
$\Omega$ to obtain pressure from (\ref{prsu}). We can then expand the 
scaled pressure at a given temperature in a Taylor series for the two 
chemical potentials $\mu_0$ and $\mu_I$,

\beqa
{P(T,\mu_0,\mu_I) \over T^4}= 
\sum^{\infty}_{n=0} \sum^{n}_{j=0} {n ! \over j ! (n-j) !} c^{jk}_n(T)
\left({\mu_0 \over T }\right)^j \left({\mu_I \over T }\right)^k ~~~;~ k=n-j,
\label{tay}
\eeqa
where,
\beqa
c^{jk}_n(T) = 
{1 \over n!} {\partial^n \left ({P(T,\mu_0,\mu_I) / T^4} \right ) 
 \over 
\partial \left({\mu_0 \over T }\right)^j \partial \left({\mu_I \over T }\right)^k}
\Big|_{\mu_0=0,\mu_I=0} ~~~.
\label{taycoff}
\eeqa
The $n= {\rm odd}$ terms vanish due to CP symmetry. Even for the 
$n= {\rm even}$ terms, due to flavour degeneracy all the coefficients 
$c^{jk}_n$ with $j$ and $k$ both odd vanish identically.
In this work we evaluate all the 10 nonzero coefficients (including
the pressure at $\mu_0 = \mu_I = 0$) upto order 
$n=6$. Some of these coefficients have already been measured on 
the LQCD \cite{sixx,eight}.
In our earlier work \cite{pnjl3}, we compared the 4 coefficients 
for $\mu_I=0$ with those of the LQCD data using improved actions
\cite{sixx}. Here we shall be able to compare 3 more coefficients
with LQCD data and also predict the behaviour of the other
3 coefficients.

Let us now identify the coefficients which we shall compare with
the LQCD data. The first set is given by,
\beqa
c_n(T) &=& {1 \over n!} \left. {\partial^n \left ({P(T,\mu_0) / T^4}
\right ) \over \partial \left({\mu_0 \over T
}\right)^n}\right|_{\mu_0=0} = c^{n0}_n~~~.
\eeqa
These coefficients were already computed upto $8$-th order and 
compared to LQCD data to $6$-th order in \cite{pnjl3}. The
new set of coefficients to be compared with the LQCD data
upto $n=6$ are,
\beqa
c^I_n(T) &=& \left. {{1 \over n!} {\partial^n \left ({P(T,\mu_0,\mu_I) /
T^4} \right ) \over 
\partial \left({\mu_0 \over T }\right)^{n-2}
\partial \left({\mu_I \over T }\right)^2 
}}\right|_{\mu_0=0,\mu_I=0} = c^{(n-2) 2}_n~~~; ~ n > 1.
\eeqa
The remaining coefficients we obtain are $c^{04}_4$, $c^{24}_6$ and
$c^{06}_6$.

To complete the comparison with the LQCD data we have looked at the
flavour diagonal ($c_n^{uu}$) and flavour off-diagonal ($c_n^{ud}$)
susceptibilities defined as,
\beqa
c^{uu}_n = {c^{n0}_n + c^{(n-2) 2}_n \over 4}, \qquad{\rm and}\qquad
c^{ud}_n = {c^{n0}_n - c^{(n-2) 2}_n \over 4} .
\eeqa
The $2$-nd order flavour diagonal and off-diagonal susceptibilities are
given by,
\beqa
  \nonumber
  \frac{\chi_{uu}(T,\mu_u=0,\mu_d=0)}{T^2} &=& \l. \frac{\partial^2
  P(T,\mu_u,\mu_d)}{\partial\mu_u^2} \r|_{{\mu_u=\mu_d=0}} = 2c_2^{uu},
  \qquad{\rm and}\qquad \\ \nonumber
  \frac{\chi_{ud}(T,\mu_u=0,\mu_d=0)}{T^2} &=& \l. \frac{\partial^2
  P(T,\mu_u,\mu_d)}{\partial\mu_u\partial\mu_d} \r|_{{\mu_u=\mu_d=0}} = 2c_2^{ud} .
\eeqa

In this work we have computed all the coefficients using the following
method. First the pressure is obtained as a function of $\mu_0$ and
$\mu_I$ for each value of $T$, and then fitted to a sixth order 
polynomial in $\mu_0$ and $\mu_I$. The quark number 
susceptibility, isospin number susceptibility and all 
other higher order derivatives are then obtained from the 
coefficients of the polynomial extracted from the fit.
In the fits we have used only the even order terms.

\subsection{Specific heat and speed of sound }
\label{susc.cvcs}

Given the thermodynamic potential $\Omega$, the energy density 
$\epsilon$ is obtained from the relation,

\beqa
\epsilon 
   = - T^2 \left . {\partial (\Omega/T) \over \partial T} \right |_V
   = - T \left . {\partial \Omega \over \partial T} \right |_V 
              + \Omega ~~~.
\eeqa

\noindent
The rate of change of energy density $\epsilon$ with temperature 
at constant volume is the specific heat $C_V$ which is given as,

\beqa
C_V = \left . {\partial \epsilon \over \partial T} \right |_V 
    = - \left . T {\partial^2 \Omega \over \partial T^2} \right |_V ~~~.
\label{sph}
\eeqa

\noindent
For a continuous phase transition one expects a divergence in $C_V$,
which, as discussed earlier, will translate into highly enhanced
transverse momentum fluctuations or highly suppressed temperature
fluctuations if the dynamics in relativistic heavy-ion collisions is
such that the system passes close to the critical end point (CEP) in the
$T-\mu_B$ plane. 

The square of velocity of sound at constant entropy $S$ is given by,

\beqa
v_s^2 = \left . {\partial P \over \partial \epsilon} \right |_S 
      = \left . {\partial P \over \partial T} \right |_V \left /
        \left . {\partial \epsilon \over \partial T} \right |_V \right .
      = \left . {\partial \Omega \over \partial T} \right |_V \left /
        \left . T {\partial^2 \Omega \over \partial T^2} \right |_V 
        \right .  ~~~.
\label{sps}
\eeqa

\noindent
Since the denominator is nothing but the $C_V$, a divergence in specific
heat would mean the velocity of sound going to zero at the CEP. 

Given the relations Eqn.\ (\ref{sph}) and Eqn.\ (\ref{sps}), we first
obtain the $\Omega(T,\mu_0=0)$ from the PNJL model. We then obtain the
derivatives using the standard finite difference method. To get points
close enough we have used cubic spline interpolations. This procedure
has been repeated for various values of $\mu_0$ and $\mu_I$.

\section{Results} \label{sc.results}

\subsection{Taylor expansion of Pressure}

As discussed in section \ref{sc.tayexp}, we extract the Taylor expansion
coefficients by fitting the pressure as a function of $\mu_0$ and
$\mu_I$ at each temperature. Data for pressure was obtained in the range
$0 < \mu_0 < 50\, {\rm MeV}$ and $0 < \mu_I < 50\, {\rm MeV}$ at all the
temperatures. Spacing between consecutive data was kept at $0.1\, {\rm
MeV}$.  We obtain all possible coefficients upto $6^{th}$ order using
gnuplot  \footnote{see http://www.gnuplot.info/} program.  The
least-squares of all the fits came out to
be $10^{-14}$ or less.  This method was already used in our earlier work
\cite{pnjl3} where we checked the reliability of such fits. Here again
we have reproduced all those coefficients satisfactorily. We shall 
first discuss the results with the standard flavour mixing in
the NJL model parameterization (\ie with $G_1 = G_2$), and then discuss 
the results for minimal ($G_2 = 0$) and maximal ($G_1 = 0$) flavour mixing.

\subsubsection{$G_1 = G_2$}

%%%%%%%%%%%%%%%%%%%%%%%%%%%%%%%%%%%%%%%%%%%%%%%%%%%%%%%%%%%%%%%%%%%%%
\begin{figure}[!tbh]
   {\includegraphics [scale=0.6] {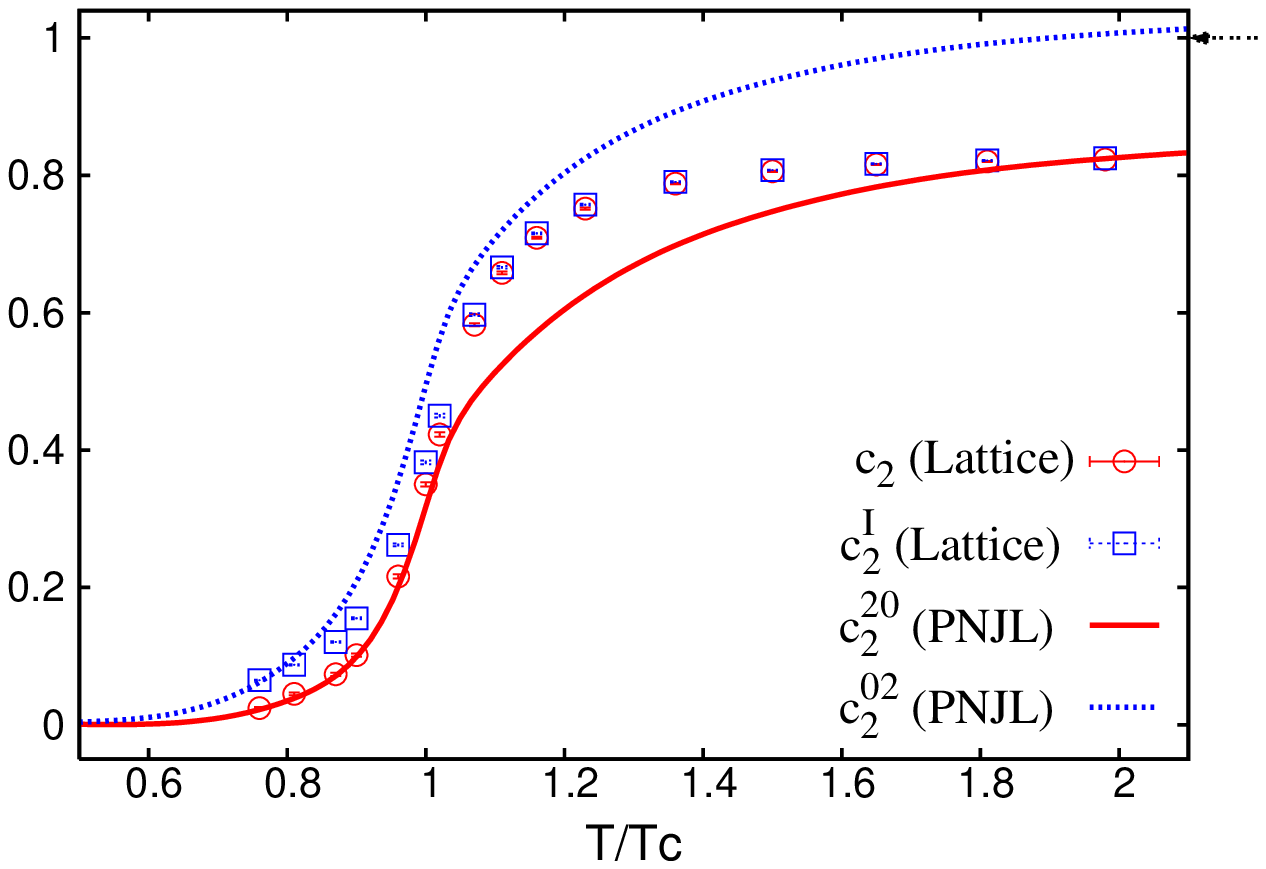}}
\hskip 0.1 in
   {\includegraphics [scale=0.6] {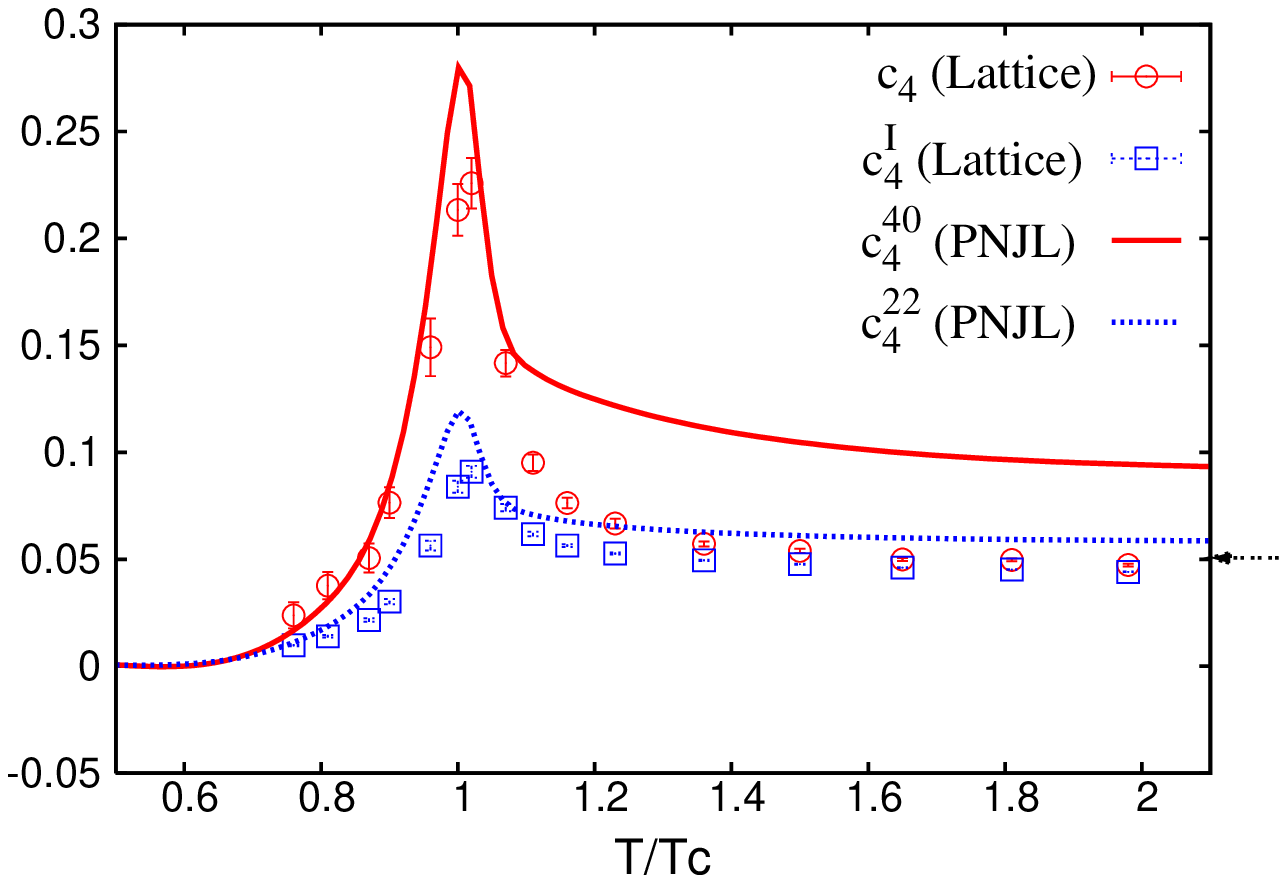}}
\vskip 0.1 in
   {\includegraphics [scale=0.6] {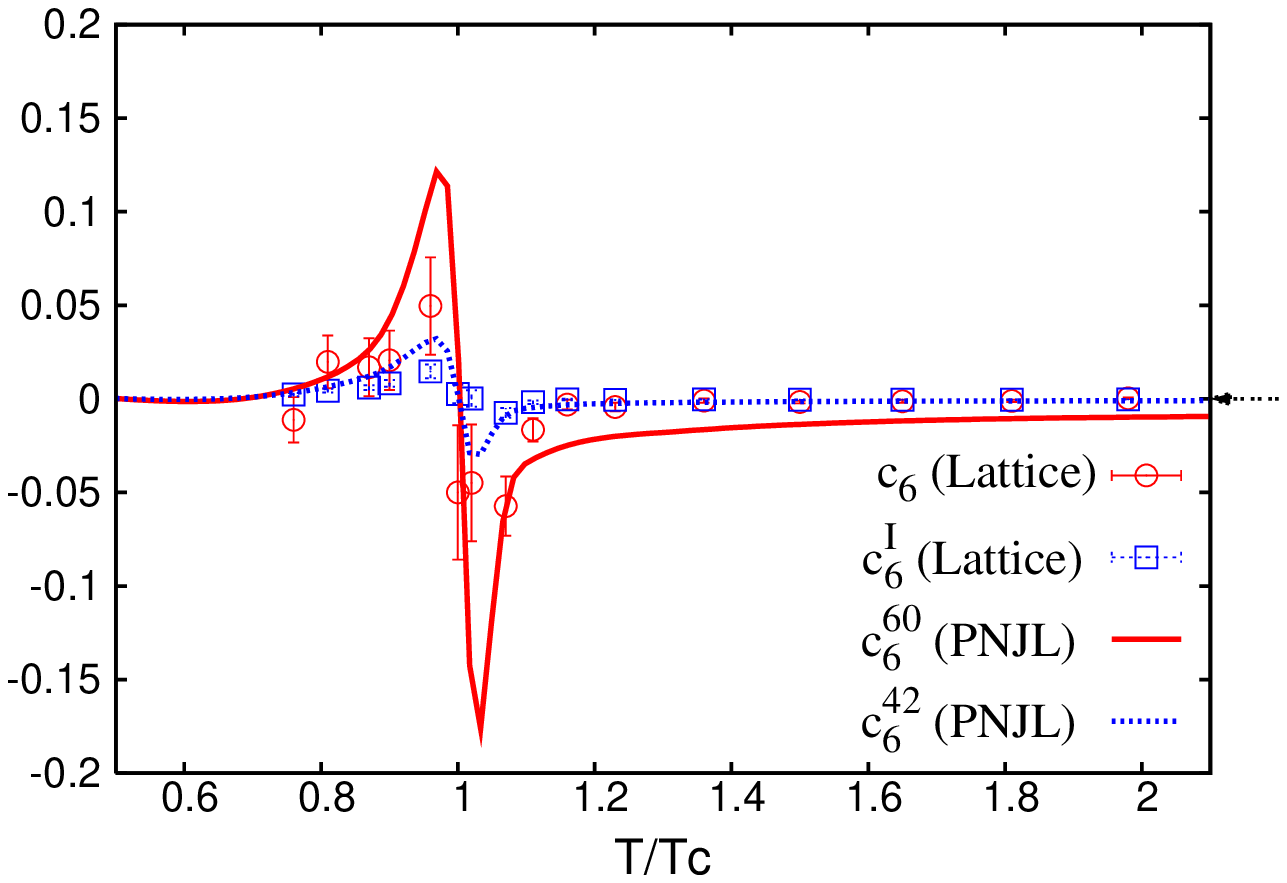}}
   \caption{The QNS and INS as a functions of $T/T_c$. Symbols are 
            LQCD data \cite{sixx}. Arrows on the right indicate the
            corresponding ideal gas values.
      }
\label{fg.scnord}\end{figure}
%%%%%%%%%%%%%%%%%%%%%%%%%%%%%%%%%%%%%%%%%%%%%%%%%%%%%%%%%%%%%%%%%%%%%

We start by presenting our results for the PNJL model with the standard
NJL Lagrangian, \ie, $G_1=G_2=G_0/2$. Note that this is the case
studied in the PNJL models of Refs.\ \cite{pnjl1,pnjl2,pnjl3}, but
without the isospin chemical potential.

We present the QNS, INS and their higher order derivatives with respect
to $\mu_0$ in Fig.\ \ref{fg.scnord}. We have also plotted the LQCD data
from Ref.\ \cite{sixx} for quantitative comparison.  At the second order
of Taylor expansion i.e. $n=2$ we find (also observed earlier in
\cite{pnjl3}) that the QNS $c_2$ compares well with the LQCD data. On
the other hand, the INS $c^I_2$ quickly reaches its ideal gas value
above $T_c$ (around $2T_c$) in our model calculations, whereas the LQCD
value are lower and matches with the value of $c_2$. 
Note that in the present form of the model the Polyakov loop itself
rises a little above 1 and saturates. This leads to the INS to rise
slightly above 1 at high temperatures. At the $4^{th}$
order we see that the values of $c_4$ (also observed in \cite{pnjl3}) in
the PNJL model matches closely with those of LQCD data for upto $T \sim
1.05 T_c$ and deviates significantly thereafter.  The coefficient
$c^I_4$ is close to the LQCD data for the full range of $T$ upto $2T_c$. 

Earlier expectation \cite{pnjl3,ratti3} was that, the mean field
analysis may not be sufficient and hence the higher order coefficient
$c_4$ in the PNJL model shows significant departure from lattice
results. This should have also meant that the INS $c^I_2$ should
be more closer to LQCD data than $c^I_4$. However, our results 
show that the INS $c^I_2$ is significantly different from the LQCD 
data above $T_c$, but $c^I_4$ is quite consistent. 
Further, we see from Fig.\ \ref{fg.scnord} that both 
the $6^{th}$ order coefficients $c_6$ and $c^I_6$ are quite consistent 
with the LQCD results. We now give a qualitative explanation 
for the PNJL results and try to understand the behaviour of the
coefficients above $T_c$. 

%%%%%%%%%%%%%%%%%%%%%%%%%%%%%%%%%%%%%%%%%%%%%%%%%%%%%%%%%%%%%%%%%%%%%
\begin{figure}[!tbh]
   {\includegraphics [scale=0.6] {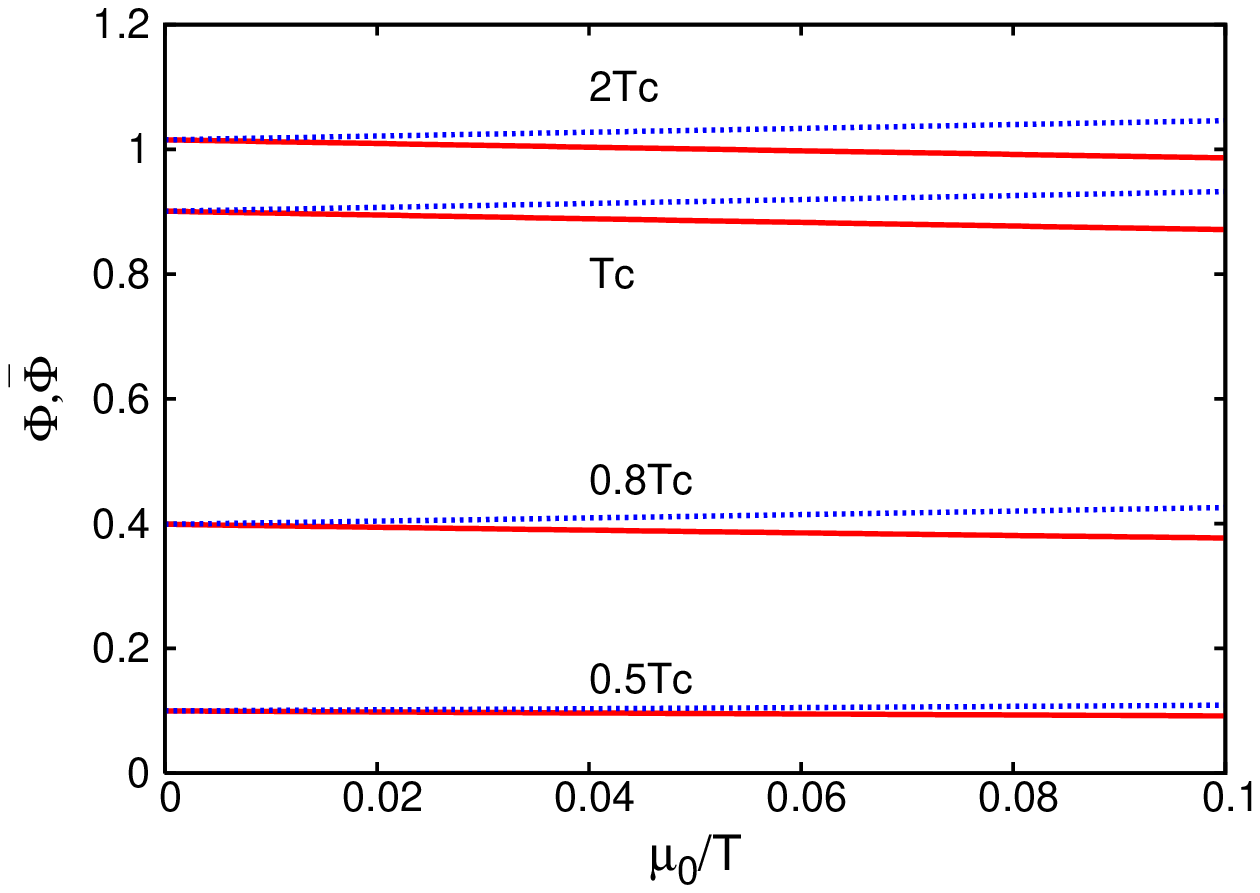}}
\hskip 0.15 in
   {\includegraphics[scale=0.6]{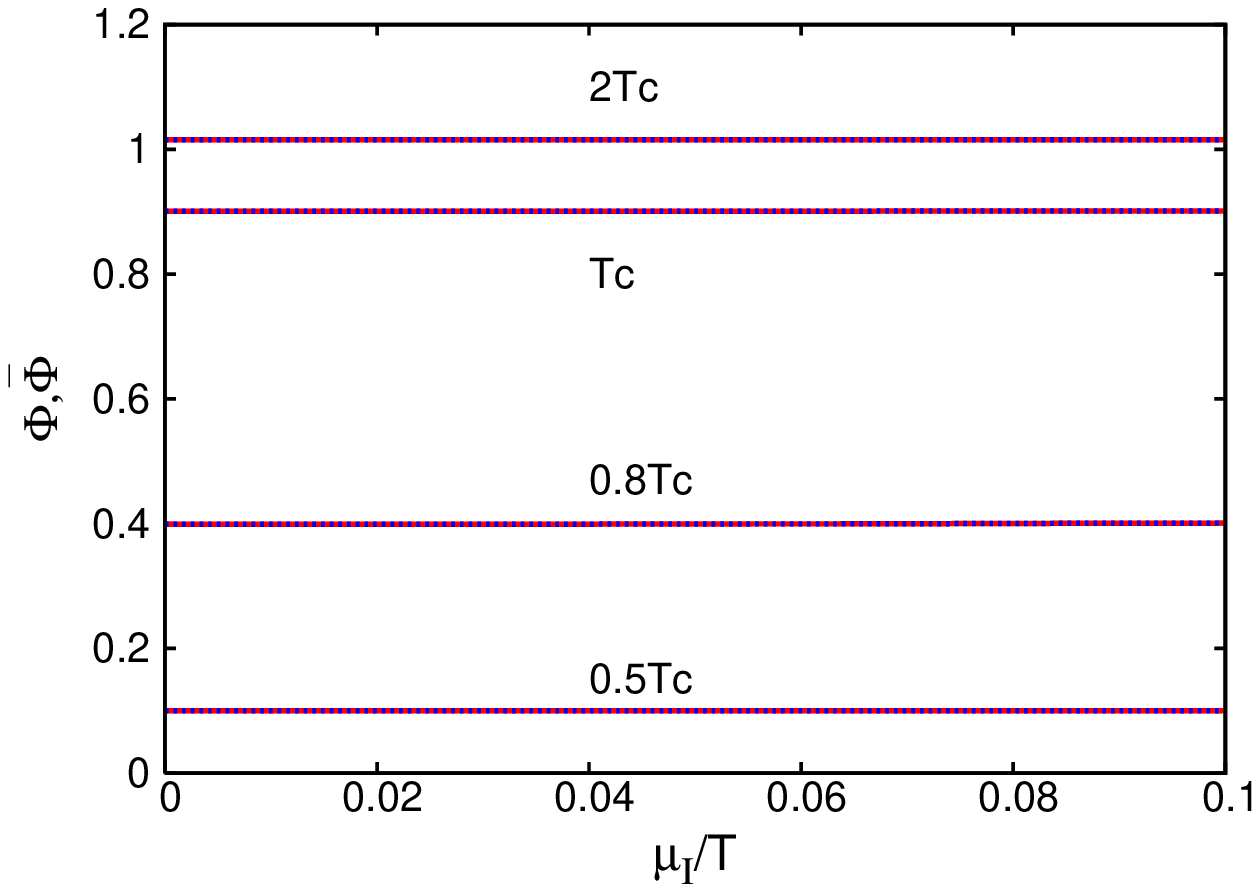}}
   \caption{{\it Left panel}: $\Phi$ (solid lines) decreases and
   $\bar{\Phi}$ (dotted lines) increases as a function of $\mu_0/T$
   ($\mu_I=0$). {\it Right panel}: $\Phi$ (solid lines) and
   $\bar{\Phi}$ (dotted lines) are equal and almost constant as a
   function of $\mu_I/T$ ($\mu_0=0$). } 
\label{fg.phimub}\end{figure}
%%%%%%%%%%%%%%%%%%%%%%%%%%%%%%%%%%%%%%%%%%%%%%%%%%%%%%%%%%%%%%%%%%%%%

We pointed out in Section\ \ref{sc.pnjl} that in the thermodynamic
potential Eqn.\ (\ref{omega}), the Polyakov loop couples to $\mu_0$ and
its conjugate couples to $-\mu_0$ due to CP symmetry. As observed in
$SU(N)$ matrix model \cite{matrix1} and also in the PNJL model
\cite{pnjl2,pnjl3}, this difference in coupling leads to splitting of
the Polyakov loop and its conjugate for any nonzero $\mu_0$. Thus even
at high temperatures when the Polyakov loop is close to $1$, it
decreases with increasing $\mu_0$ and its conjugate increases (see left
panel of Fig. \ref{fg.phimub}).  This means that the $\mu_0$ dependence
of pressure is not the same as that for an ideal gas. Hence the
coefficients $c_2$ and $c_4$ are both quite off from their respective
ideal gas values. Also note that though $c_6$ is close enough, it is
still distinctly different from zero.  On the other hand for $\mu_0=0$,
the Polyakov loop as well as its conjugate couples to both the $\mu_I$
and $-\mu_I$.  They are, thus, equal (see right panel of Fig.
\ref{fg.phimub}), and also found to be almost constant for small
$\mu_I$. So the temperature dependence of the INS and its $\mu_I$
derivatives should reach the ideal gas behaviour above $T_c$.  For the
coefficients which are mixed derivatives of $\mu_0$ and $\mu_I$ the
behaviour should be somewhere in between. And indeed we see that
$c^I_2$, $c^I_4$ and $c^I_6$ in Fig.\ \ref{fg.scnord} are quite close to
their respective ideal gas values above $T_c$. Thus the LQCD results 
that show almost equal values of QNS and INS, indicate that the splitting
between the Polyakov Loop and its conjugate in the $\mu_I=0$ direction
for $T > 1.5 T_c$ is almost negligible (also supported by pQCD). 
This splitting was taken to be absolutely zero in the recent report 
with the PNJL model in Ref. \cite{ratti4}.

%%%%%%%%%%%%%%%%%%%%%%%%%%%%%%%%%%%%%%%%%%%%%%%%%%%%%%%%%%%%%%%%%%%%%
\begin{figure}[!tbh]
   {\includegraphics [scale=0.45] {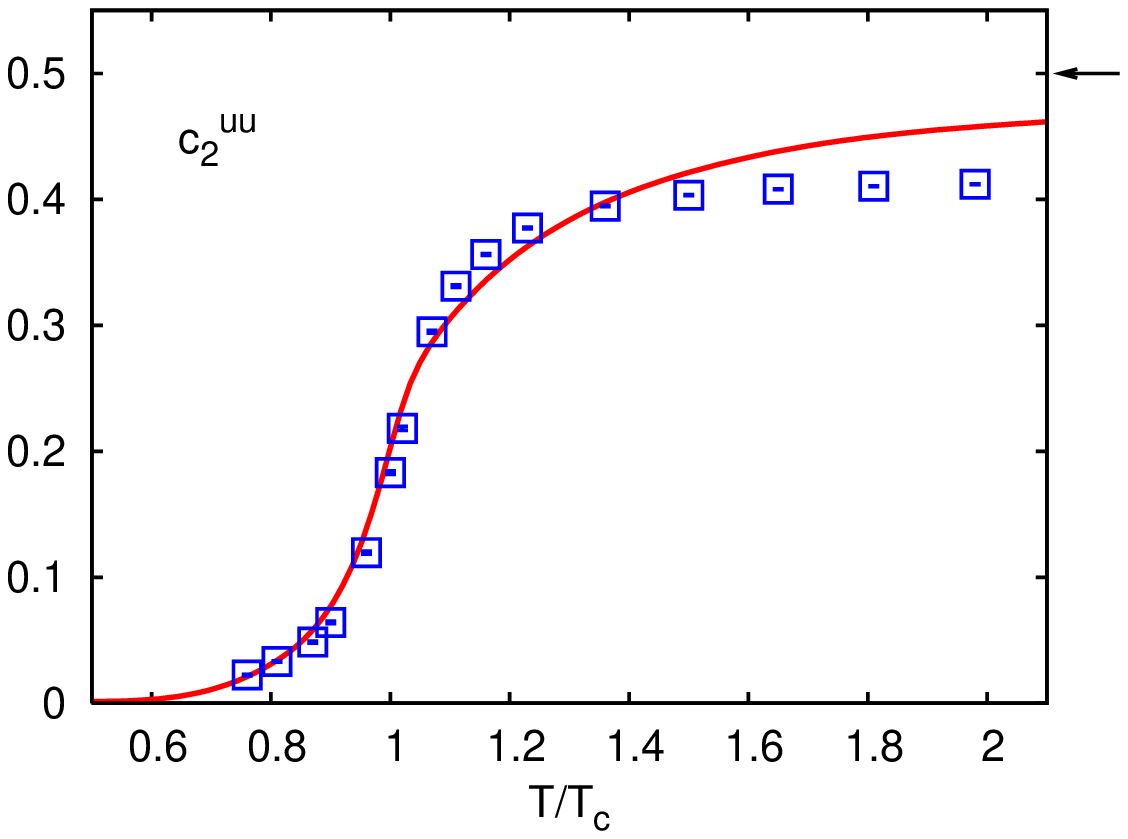}}
   {\includegraphics[scale=0.45]{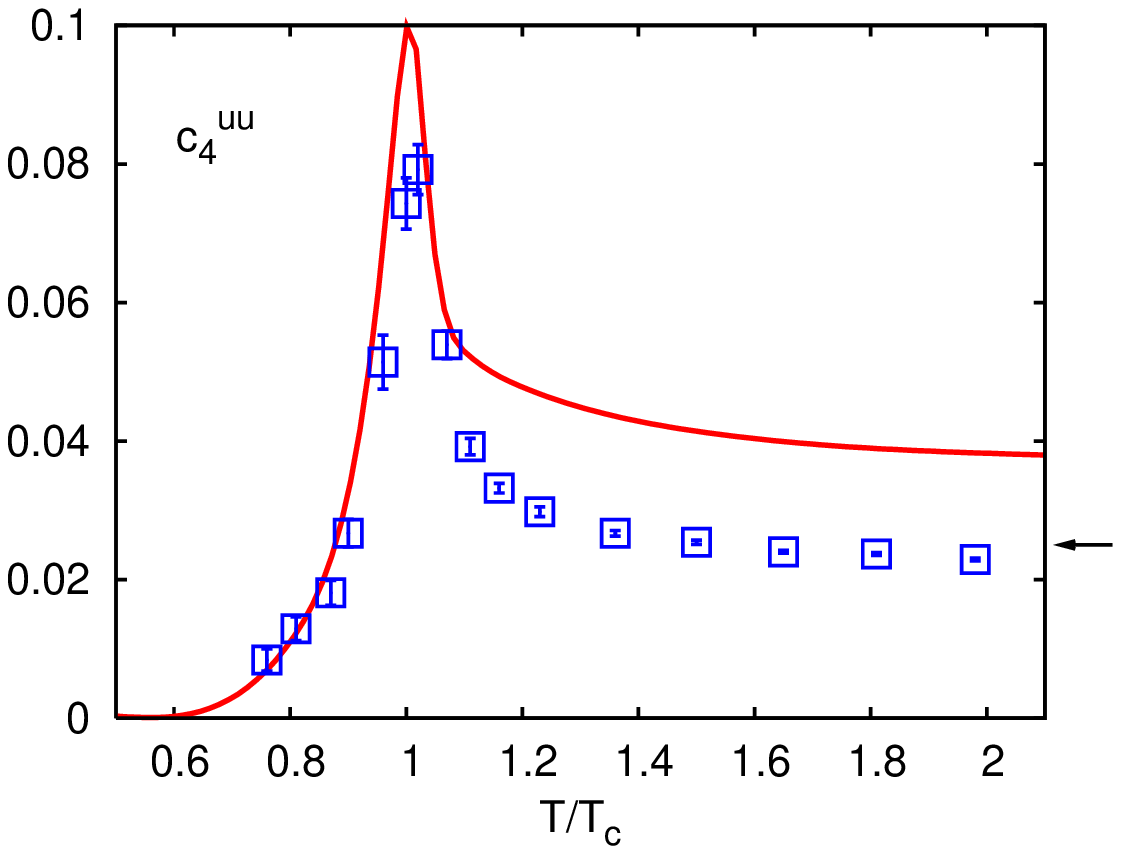}}
   {\includegraphics[scale=0.45]{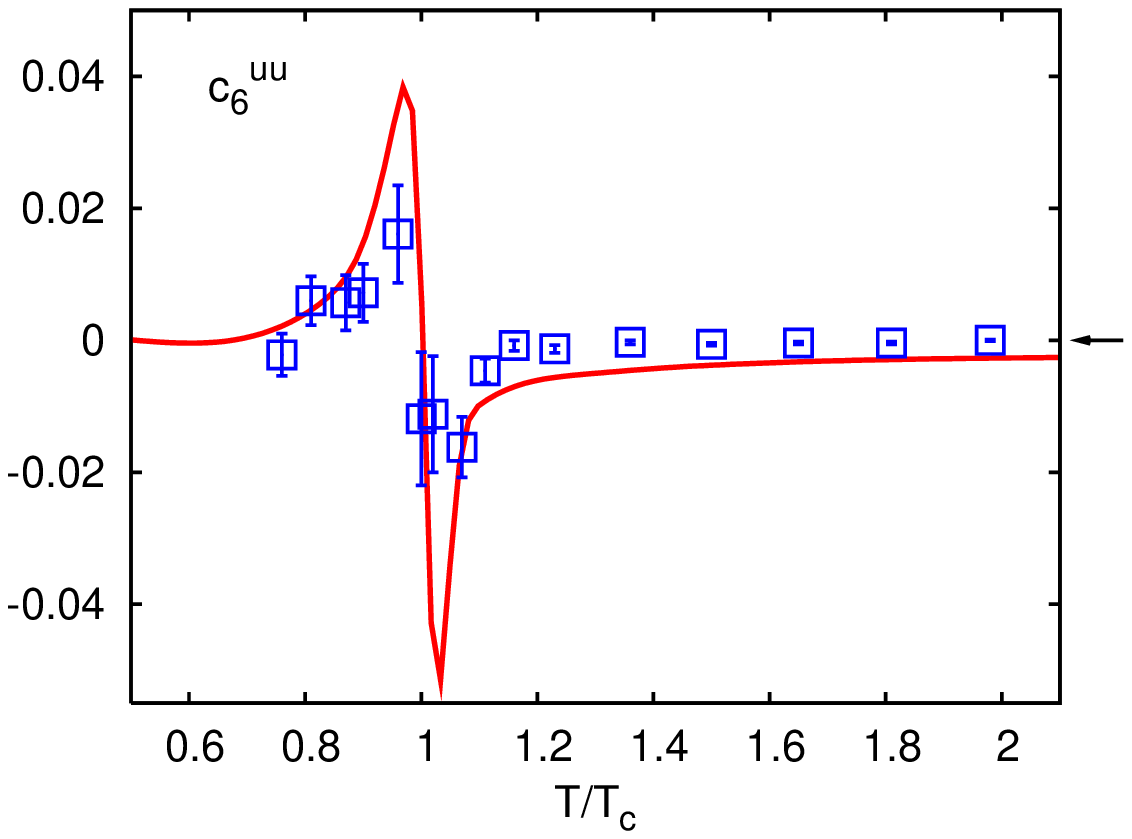}}
   {\includegraphics [scale=0.45] {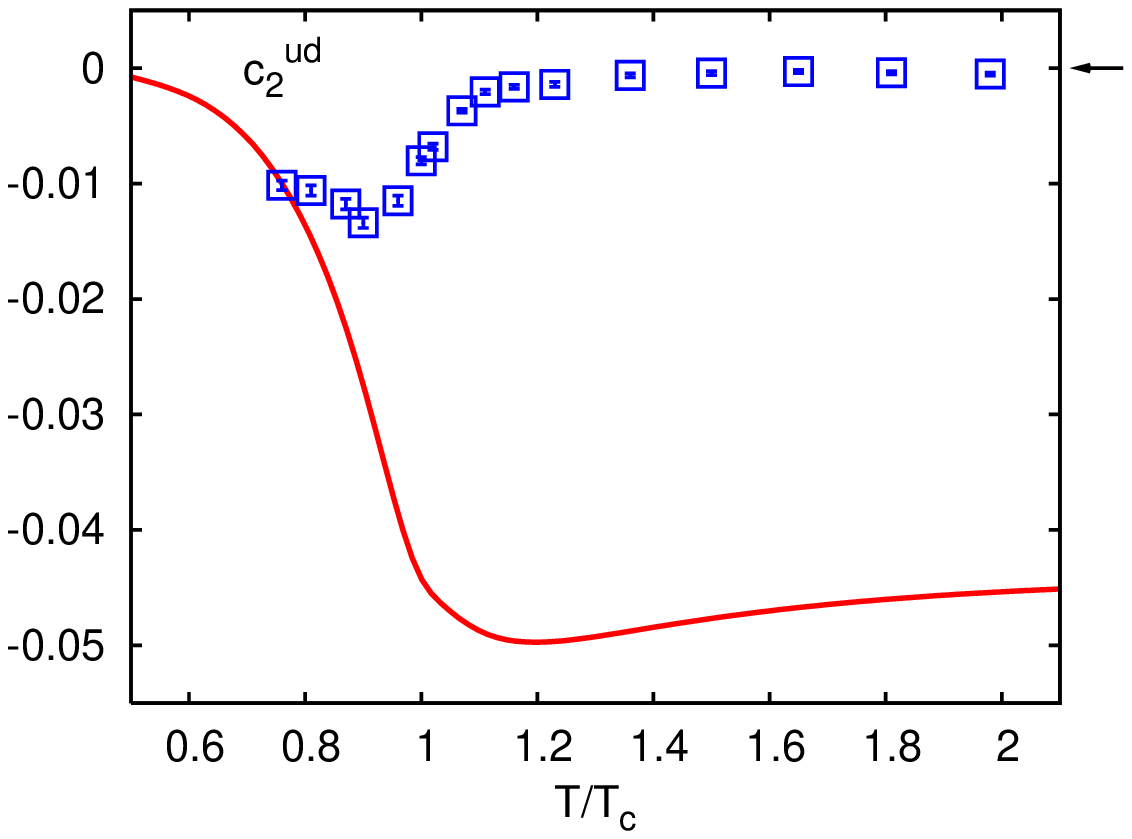}}
   {\includegraphics[scale=0.45]{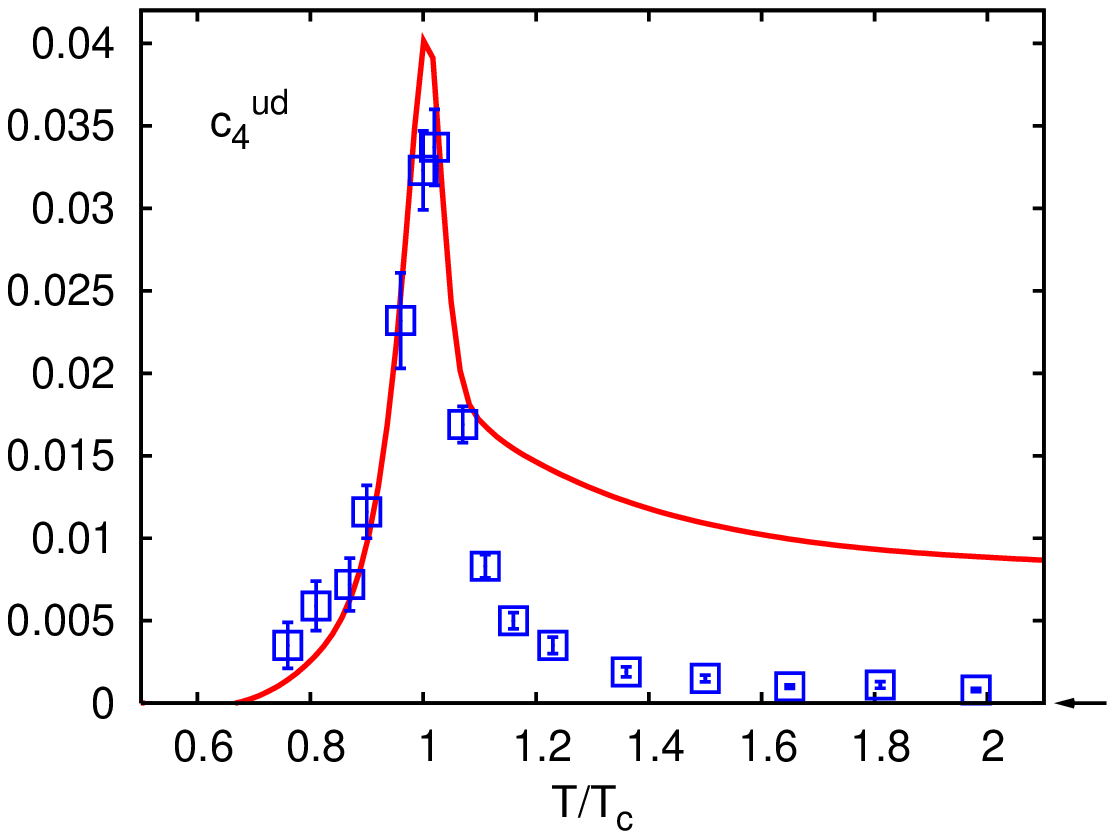}}
   {\includegraphics[scale=0.45]{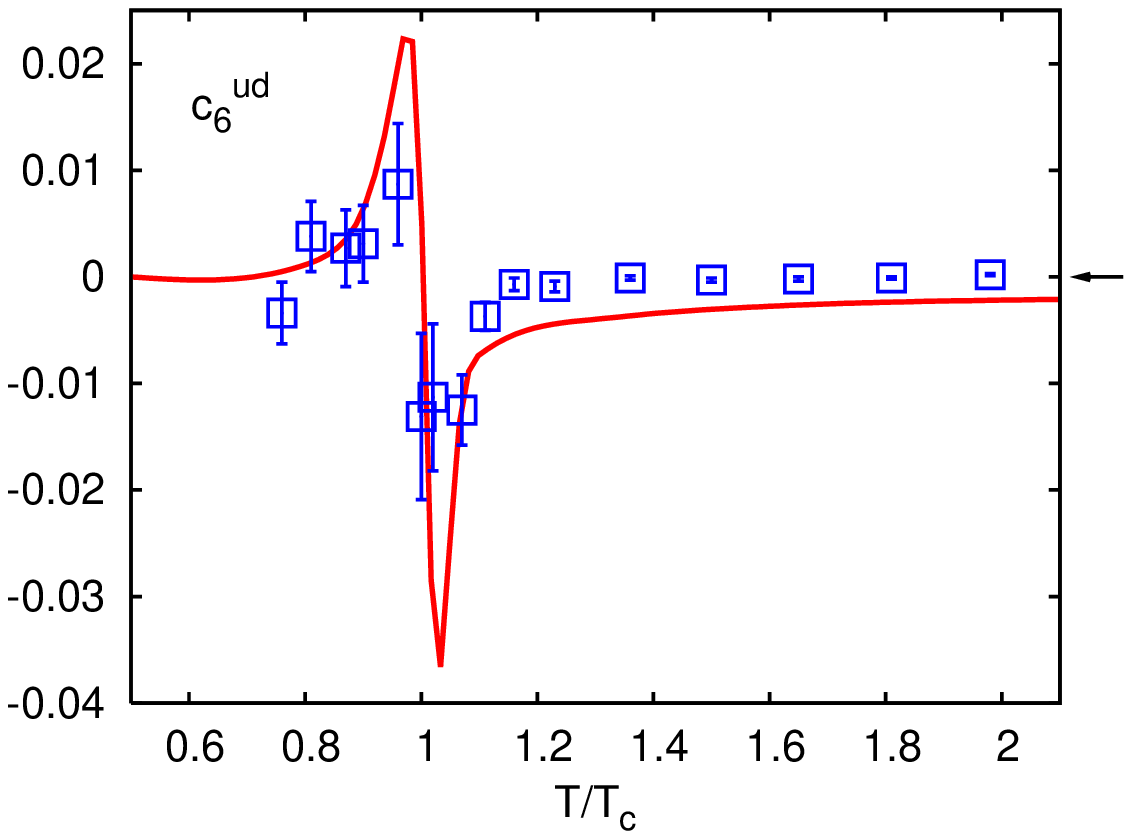}}
   \caption{The flavour diagonal ({\it upper row}) and flavour
   off-diagonal ({\it lower row}) susceptibilities for $n = 2$, $4$ and
   $6$ as functions of $T/T_c$. Symbols are LQCD data \cite{sixx}. The 
   arrows on the right indicate the respective ideal gas values.}
\label{fg.diodi}\end{figure}
%%%%%%%%%%%%%%%%%%%%%%%%%%%%%%%%%%%%%%%%%%%%%%%%%%%%%%%%%%%%%%%%%%%%%

In order to investigate these discrepancies between the results form the
PNJL model and the LQCD data more closely, we have also calculated the
flavour diagonal ($c_n^{uu}$) and off-diagonal ($c_n^{ud}$)
susceptibilities, defined in Section\ \ref{sc.tayexp}, upto $6$-th
order. These are shown in Fig.\ \ref{fg.diodi}.  Except for $c_2^{uu}$,
all the other LQCD results for flavour diagonal susceptibilities are
close to their respective ideal gas values from around $1.2T_c$ onwards.
The PNJL model values for the diagonal coefficient $c^{uu}_2$ seem to be
more or less consistent with the LQCD data. The most striking
discrepancy with the LQCD data shows up in the $2$-nd order flavour
off-diagonal susceptibility $c^{ud}_2$. As discussed earlier, $c_2^{ud}$
signifies the mixing of $u$ and $d$ quarks through the contribution of
the two disconnected $u$ and $d$ quark loops. While the LQCD data shows
that this kind of correlation between the $u$-$d$ flavours are almost
zero just away form $T_c$, the PNJL model results remains significant
even upto $2 T_c$. The negativity of $c_2^{ud}$ (see Fig.\
\ref{fg.diodi}) indicates that the dominant correlation is between 
$u$ quarks and $d$ anti-quarks and vice-verse, \ie, pion-like. Hence 
putting in the dynamical pion condensate may throw some light on this
issue. Also addition of any new couplings (e.g. as shown for the 
isoscalar-vector and isovector-vector couplings for NJL model in 
Ref. \cite{redlich1}) may have important consequences for these 
suscpetibilities.

%%%%%%%%%%%%%%%%%%%%%%%%%%%%%%%%%%%%%%%%%%%%%%%%%%%%%%%%%%%%%%%%%%%%%
\begin{figure}[!tbh]
   {\includegraphics [scale=0.6] {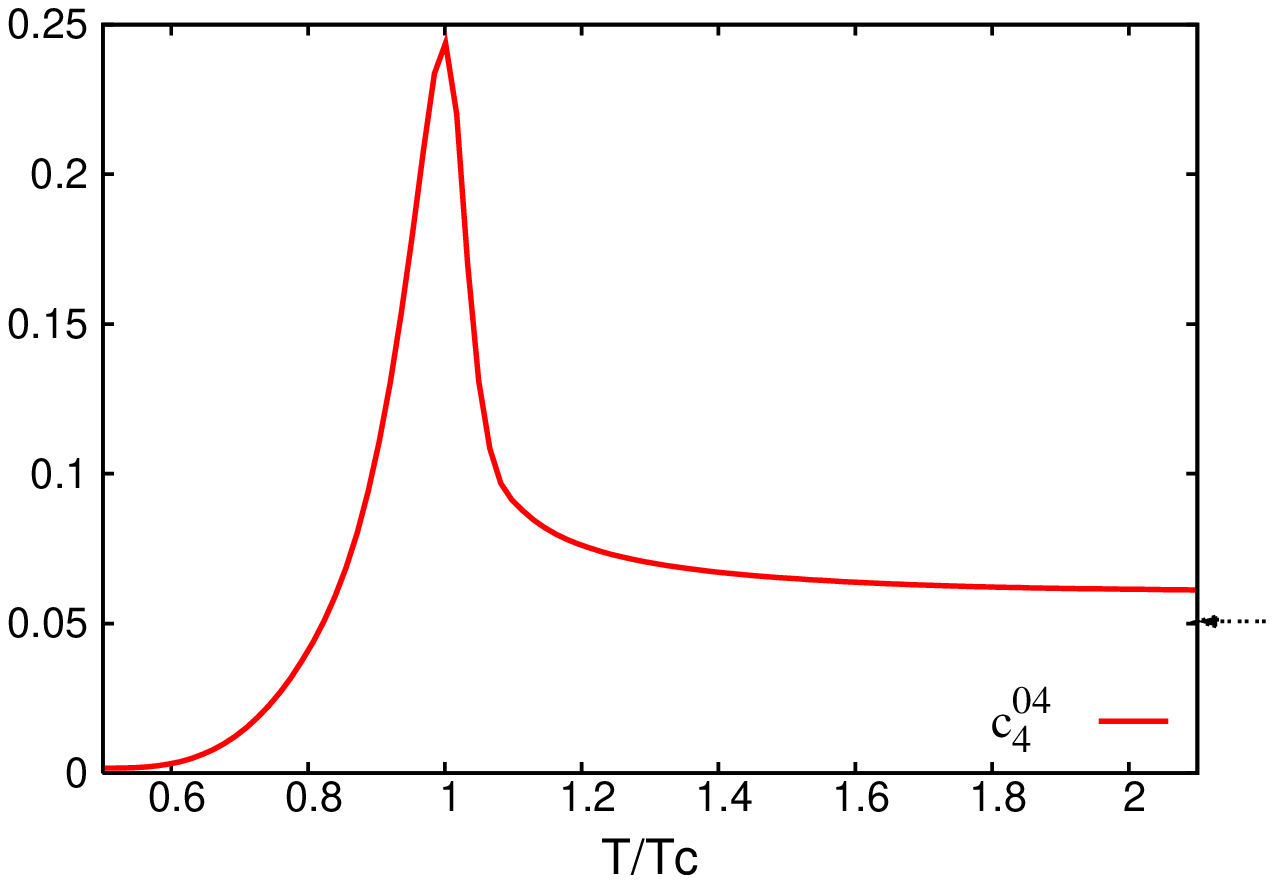}}
\hskip 0.1 in
   {\includegraphics [scale=0.6] {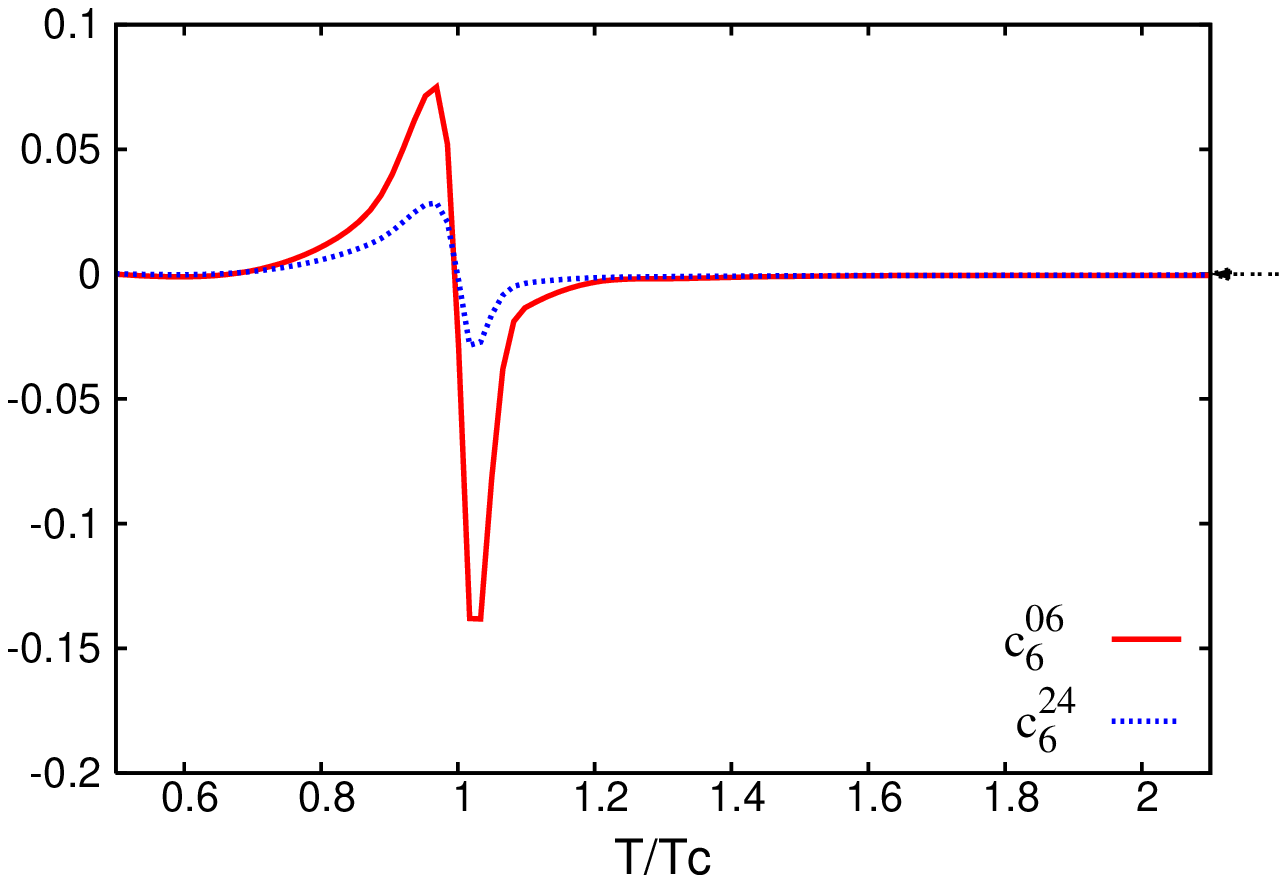}}
   \caption{$c^{04}_4$ and $c^{06}_6$ and $c^{24}_6$ as functions of 
            $T/T_c$. Arrows on the right indicate the respective ideal 
	    gas values.}
\label{fg.diag}\end{figure}
%%%%%%%%%%%%%%%%%%%%%%%%%%%%%%%%%%%%%%%%%%%%%%%%%%%%%%%%%%%%%%%%%%%%%

Again from Fig.\ \ref{fg.diodi}, the $4$-th order diagonal ($c^{uu}_4$)
as well as the off-diagonal ($c^{ud}_4$) coefficients show a behaviour
similar to $c^{40}_4$. Whereas the LQCD data reaches the ideal gas value
above $T_c$, the PNJL values are quite distinctly separated. Finally, at
the $6^{th}$ order the behaviour for both diagonal and off-diagonal
coefficients in the PNJL model and LQCD are quite consistent.

We now present the temperature dependence of the remaining nonzero
coefficients (Fig.\ \ref{fg.diag}). $c^{04}_4$ is the $4$-th order
diagonal coefficient in the isospin direction. In contrast to $c^{40}_4$
we see that $c^{04}_4$ approaches the ideal gas value quite fast above
$T_c$. The behaviour of $c^{06}_6$ is quite similar to its counterpart
$c^{60}$. This is in accordance to the expectation, as discussed
earlier. Same is true for the coefficient $c^{24}_6$.

\subsubsection{$G_1 \ne G_2$}

%%%%%%%%%%%%%%%%%%%%%%%%%%%%%%%%%%%%%%%%%%%%%%%%%%%%%%%%%%%%%%%%%%%%%
\begin{figure}[!tbh]
   {\includegraphics [scale=0.4] {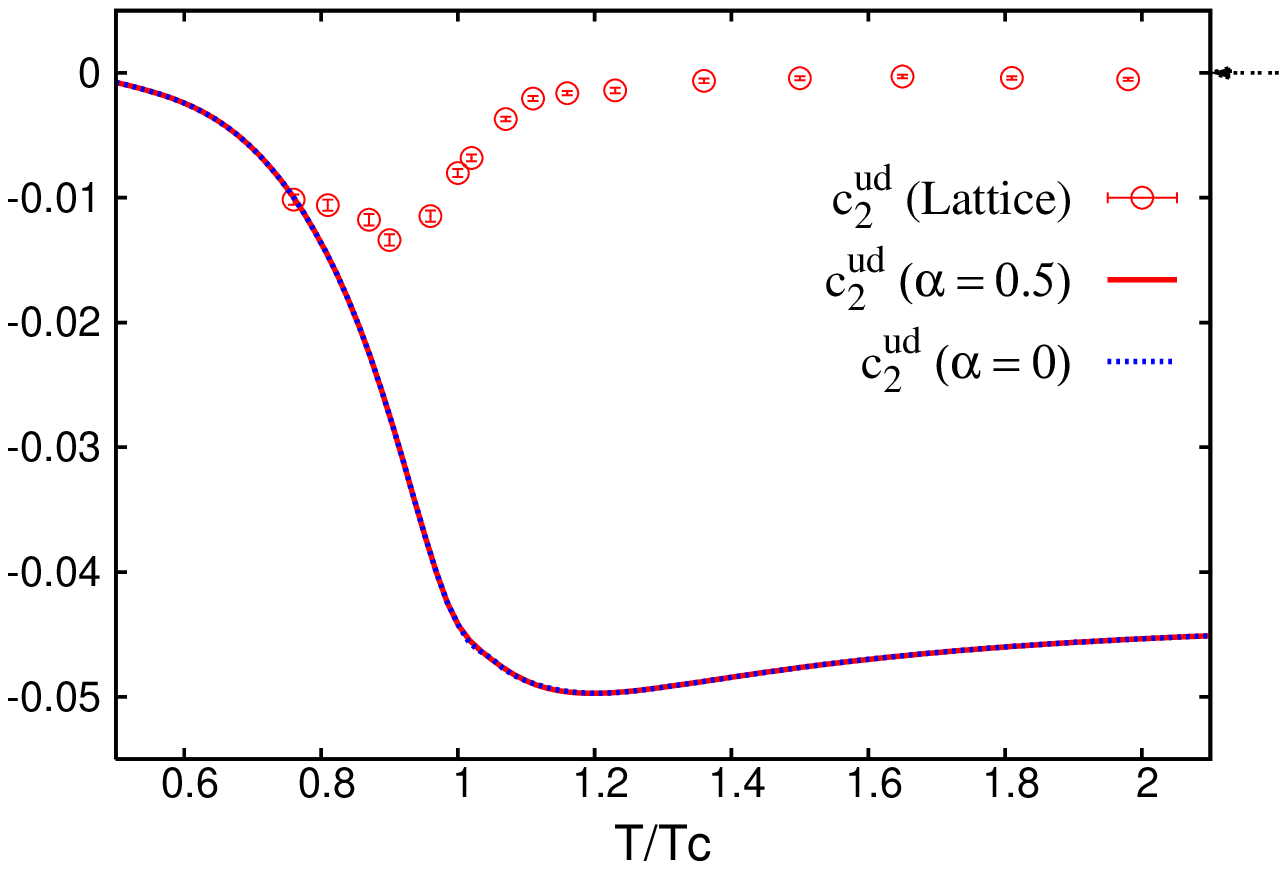}}
\hskip 0.1 in
   {\includegraphics [scale=0.4] {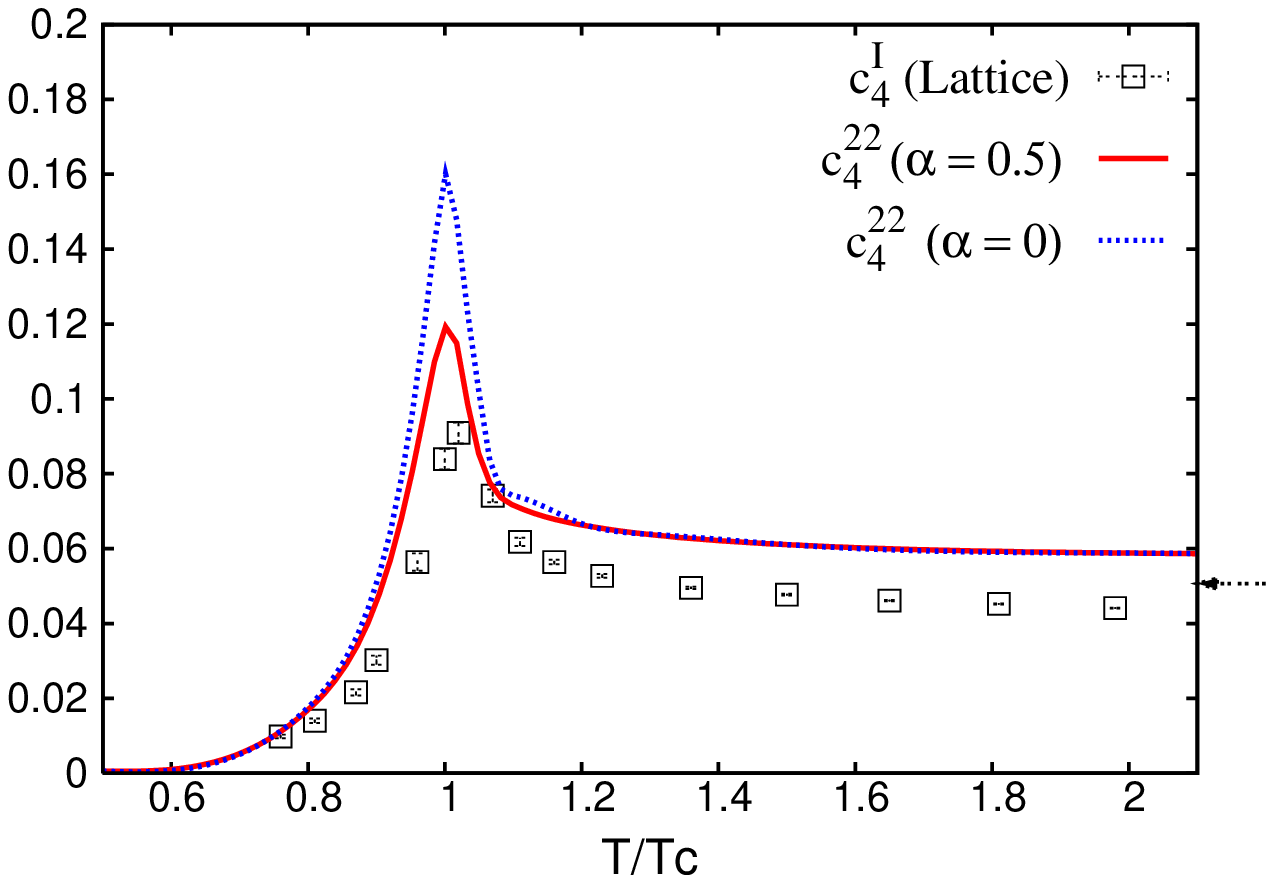}}
\hskip 0.1 in
   {\includegraphics [scale=0.4] {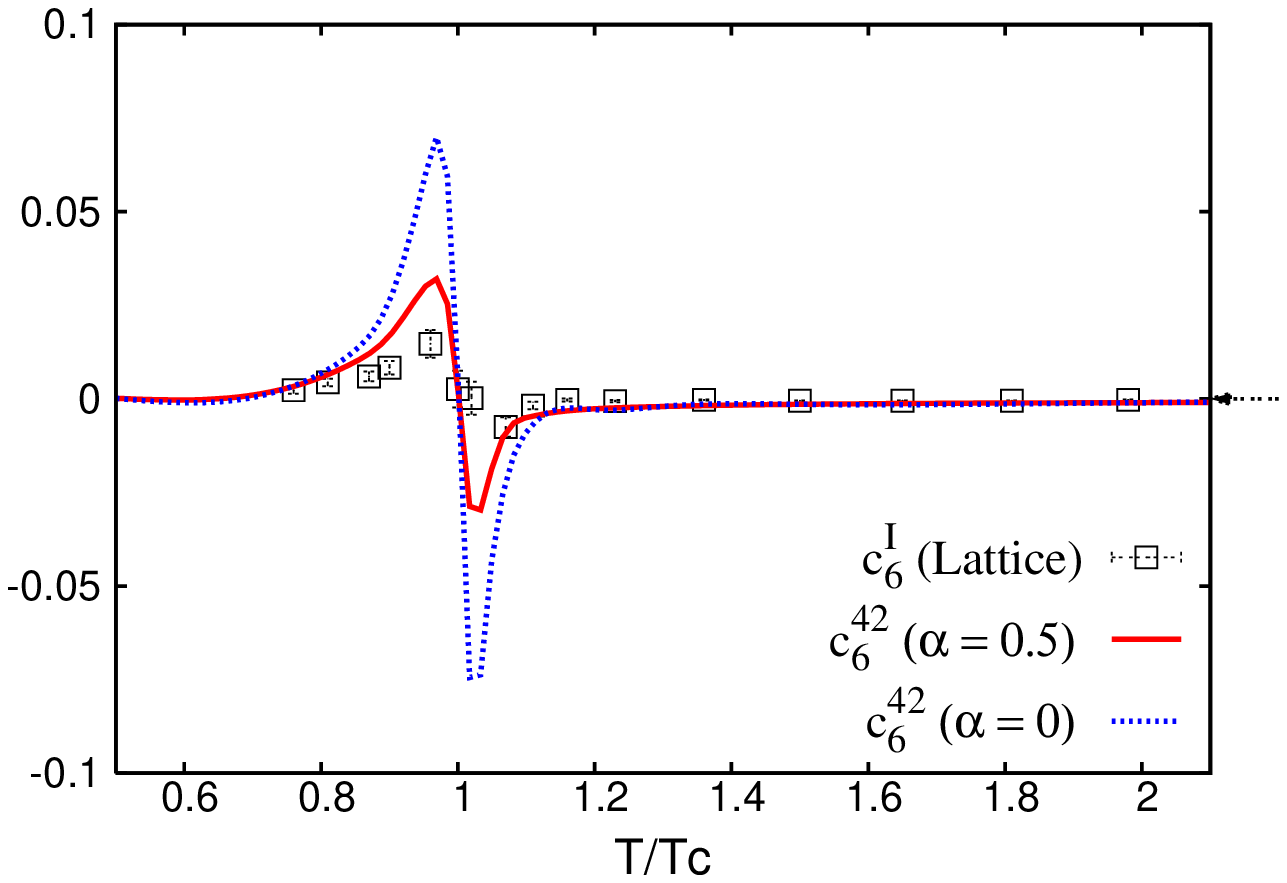}}
   \caption{{\it Left panel:} $c^{ud}_2$ is independent of $\alpha$.
   {\it Middle and Right panel:} Dependence of some off-diagonal 
   coefficients on the flavour mixing parameter $\alpha$. Symbols
   are Lattice data \cite{sixx}. Arrows on the right indicate the
   corresponding ideal gas values. 
   }
\label{fg.aldep}\end{figure}
%%%%%%%%%%%%%%%%%%%%%%%%%%%%%%%%%%%%%%%%%%%%%%%%%%%%%%%%%%%%%%%%%%%%%

Since the PNJL model has problem in reproducing the LQCD data for
$c_2^{ud}$ which is a measure of the flavour-flavour correlation, it is
interesting to have a closer look at the effect of flavour-mixing on
different susceptibilities. As discussed earlier, the parameterization
$G_1 = (1-\alpha) G_0$ and $G_2 = \alpha G_0$ enables one to tune the
instanton induced flavour mixing by varying the value of $\alpha$
between 0 and 1. Here we discuss the two extreme cases of $\alpha =1$
(maximal mixing) and $\alpha=0$ (zero mixing). We have re-calculated all
$c_n$ and $c_n^I$, upto $n=6$, for $\alpha=0,\ {\rm and},\ 1$.  We found
that all the diagonal coefficients, including $c^I_2$ and  $c_4$  whose
behaviour are the most drastically different in the PNJL model and in
LQCD, are independent of the values of $\alpha$. As a consequence, the
$2$-nd order flavour off-diagonal susceptibility
[$c_2^{ud}=(c_2^I-c_2)/4$] is also unaffected by the instanton induced
flavour-mixing effects. 

The above fact can be understood from the following reasoning.  We
mentioned in section \ref{sc.pnjl} that the quark condensates $\sigma_u$
and $\sigma_d$ are equal to each other for either of the cases $\mu_0 =
0$ and $\mu_I = 0$. This is clear from Eqn.\ \ref{eq.sig-u-d}. Now
$G_1$ and $G_2$ couple only to the $\sigma_u$ and $\sigma_d$. So for
$\sigma_u =\sigma_d$, we only get the combination $G_1 + G_2 = G_0$,
which is a constant. Thus none of the physics in the $\mu_0 = 0$ and
$\mu_I = 0$ directions depend on the value of $\alpha$, implying that
the diagonal derivatives in these two directions will also be
independent of $\alpha$.

However, the mixed derivatives can have dependence on $\alpha$. This is
because the values of $\sigma_u$ and $\sigma_d$ can be different when
both $\mu_0$ and $\mu_I$ are together nonzero. This was seen in
Ref.\ \cite{buballa} for the normal NJL model. But those authors also
found that there is a critical value of $\alpha_c \approx 0.11$ above
which the condensates $\sigma_u$ and $\sigma_d$ become equal even for
both $\mu_0$ and $\mu_I$ being nonzero. Here, for the PNJL model we have
found that all the mixed derivatives upto $6^{th}$ order are exactly
equal for the two cases $\alpha = 0.5$ (standard mixing used in NJL and
PNJL models) and $\alpha = 1$ (maximal mixing) which is in accordance to
the results of the above reference. We hope to obtain the value for
$\alpha_c$ for the PNJL model in future. For $\alpha = 0$ all the 
off-diagonal coefficients were found to differ from those at $\alpha=0.5$. 

The left-most panel in Fig. \ref{fg.aldep} shows the independence of
$c^{ud}_2$ on $\alpha$. The rest figures show one representative 
coefficient each for $n = 4,\, {\rm and}\ 6$. As can be seen, the 
instanton effects quite significantly suppress the temperature
variation of these coefficients near $T_c$.  Also it can 
be observed from Fig.\ \ref{fg.aldep}, that the LQCD data favours larger
amount of instanton induced flavour-mixing.

\subsection{Dependence of $C_V$ and $v_s^2$ on $\mu_0$ and $\mu_I$}

%%%%%%%%%%%%%%%%%%%%%%%%%%%%%%%%%%%%%%%%%%%%%%%%%%%%%%%%%%%%%%%%%%%%%
\begin{figure}[!tbh]
   {\includegraphics [scale=0.65] {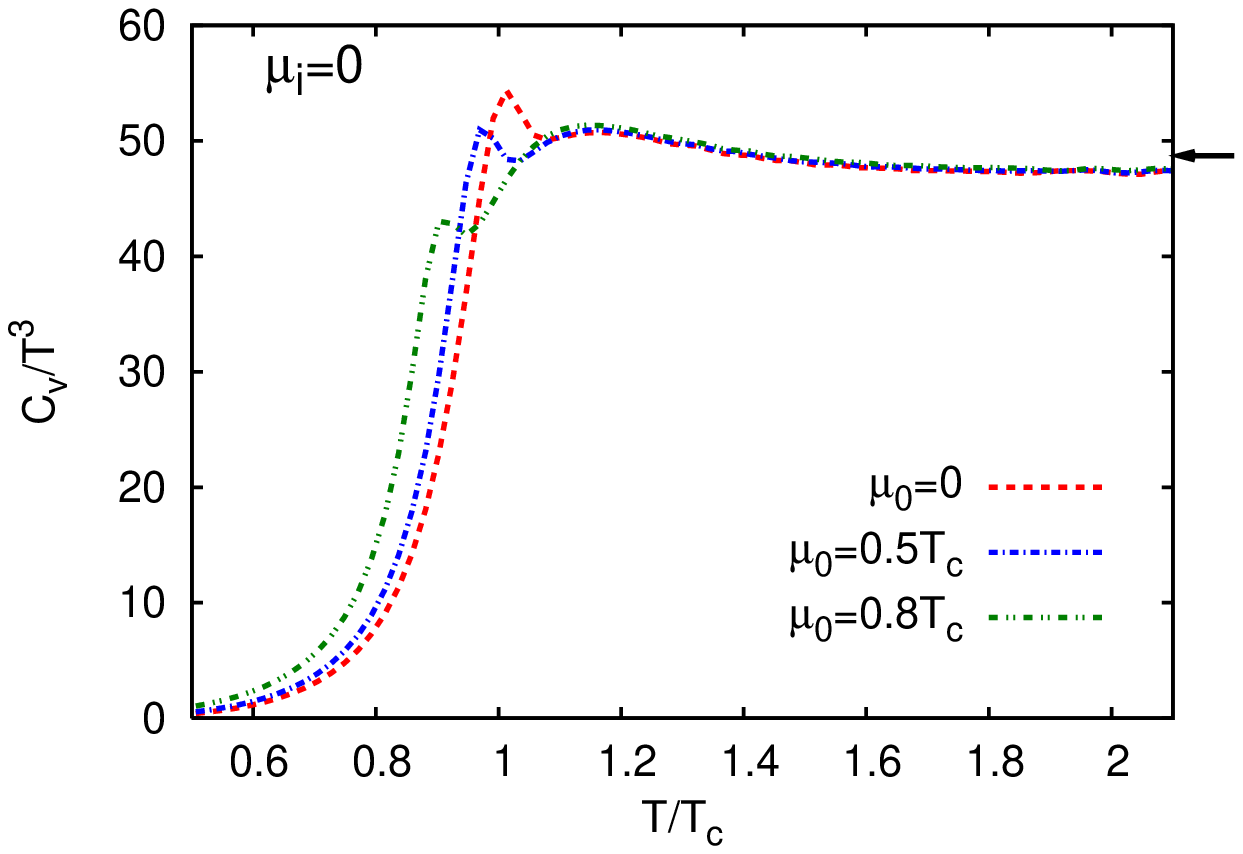}}
\hskip 0.1 in
   {\includegraphics [scale=0.65] {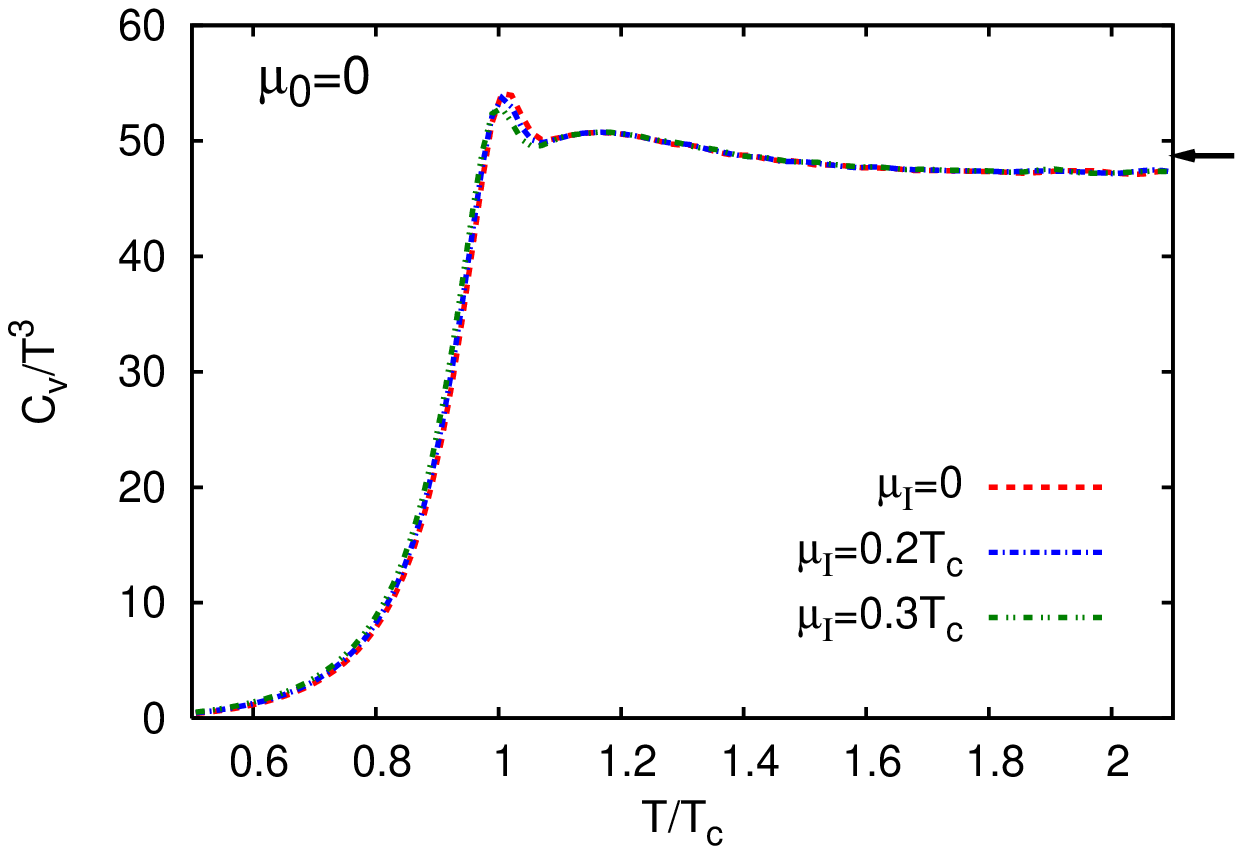}}
   \caption{$C_V$ as a function of $T/T_c$. {\it Left panel} shows the
   variation with $\mu_0$; {\it Right panel} shows the variation with
   $\mu_I$. Arrows on the right indicate the ideal gas value for
   $\mu_0=\mu_I=0$. }
\label{fg.cv}\end{figure}
%%%%%%%%%%%%%%%%%%%%%%%%%%%%%%%%%%%%%%%%%%%%%%%%%%%%%%%%%%%%%%%%%%%%%

%%%%%%%%%%%%%%%%%%%%%%%%%%%%%%%%%%%%%%%%%%%%%%%%%%%%%%%%%%%%%%%%%%%%%
\begin{figure}[!tbh]
   {\includegraphics [scale=0.65] {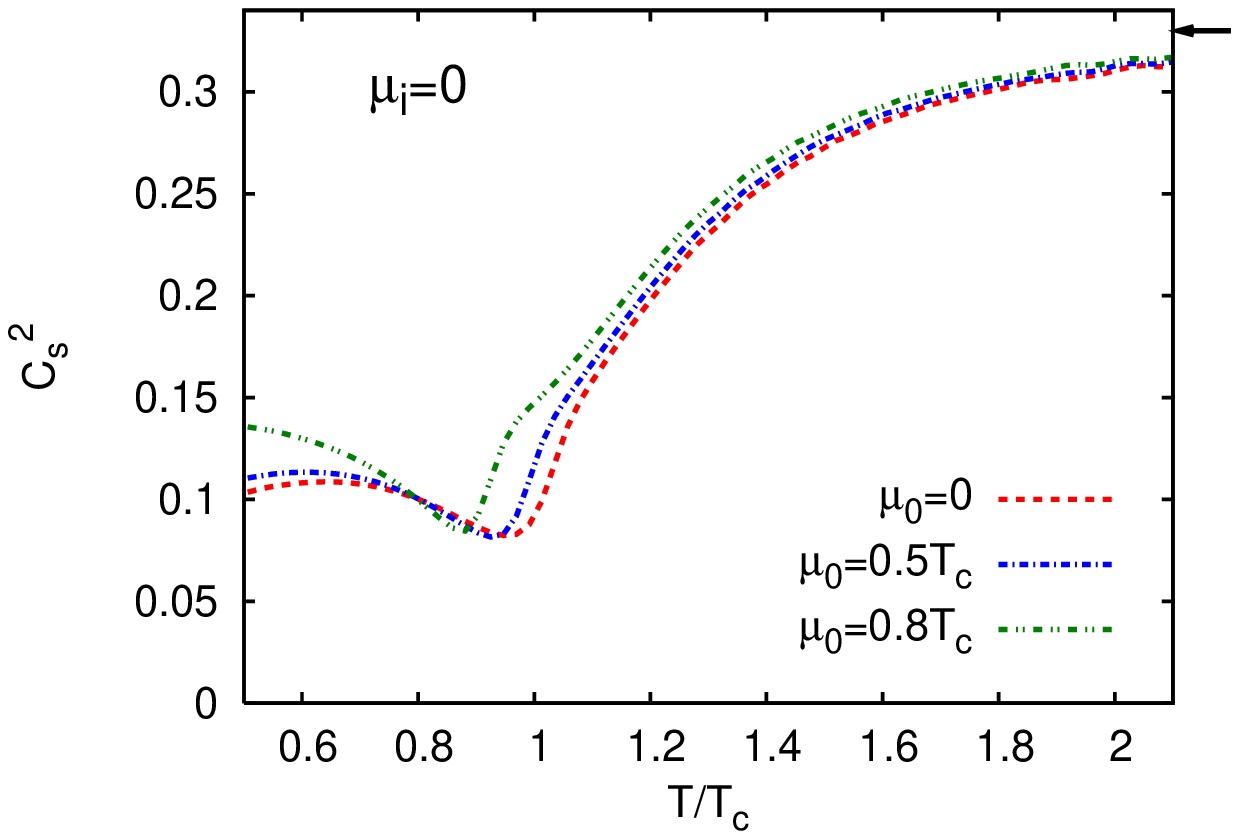}}
\hskip 0.1 in
   {\includegraphics [scale=0.65] {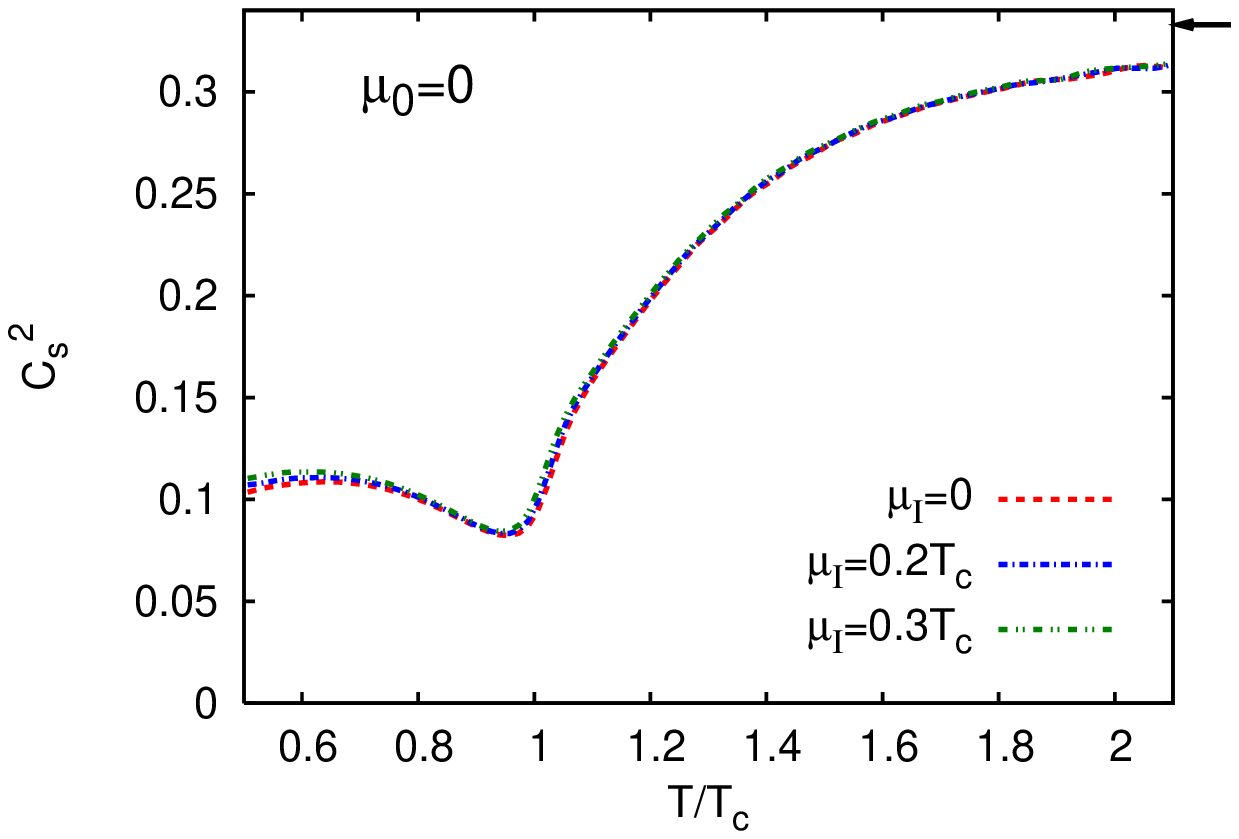}}
   \caption{$v^2_s$ as a function of $T/T_c$. {\it Left panel} shows the
   variation with $\mu_0$; {\it Right panel} shows the variation with
   $\mu_I$. Arrows on the right indicate the ideal gas value for
   $\mu_0=\mu_I=0$. }
\label{fg.cs}\end{figure}
%%%%%%%%%%%%%%%%%%%%%%%%%%%%%%%%%%%%%%%%%%%%%%%%%%%%%%%%%%%%%%%%%%%%%

Here we present the chemical potential dependence of specific heat $C_V$
and the speed of sound $v_s$. The range of the three representative
values of $\mu_0$ and $\mu_I$ are such that neither the diquark physics
nor the pion condensation becomes important.  In the ideal gas limit the
expression for $C_V$ is as a function of temperature $T$ and either of
the chemical potentials $\mu_0$ or $\mu_I$ is given by, $C_V/T^3 =
(74\pi^2/15) + 6(\mu^2_{0,I}/T^2)$.  Thus, for large temperatures and
not so large chemical potentials, it can be expected that the $C_V$ is
more or less independent of $\mu_{0,I}$'s. This is borne out in the
PNJL model as seen in Fig. \ref{fg.cv}. At low temperatures however,
there can be non-trivial contribution from chemical potential. As
illustrated in Fig.  \ref{fg.cv}, the low temperature behaviour is away
from ideal gas, but there is significant difference in the values of
$C_V$ as a function of $\mu_0$. In the range of $\mu_I$ considered, even
for $T<T_c$ there seems to no significant isospin effects. Another
interesting feature is that as a function of $\mu_0$, the peak of $C_V$
which appears at $T_c$ shifts towards lower temperatures. This signifies
that the transition temperature may decrease and also the nature of
transition may change as the chemical potentials increase. A decrease of
$T_c$ with increasing $\mu_0$ and $\mu_I$ is consistent with what have
been found on the Lattice \cite{allton,kogut1}. We hope to address this
issue through the analysis of chiral susceptibility in a future
publication.

The speed of sound in the ideal gas limit is the same $\sqrt{3}$ for any
given temperature and chemical potential. As shown in Fig.\ \ref{fg.cs}
the $v^2_s$ for different $\mu_0$ and $\mu_I$ merges towards the ideal
gas value at large temperatures. However, even above $T_c$, there is
significant increase in $v^2_s$ for increase in $\mu_0$. So for nonzero
quark matter density the speed of sound is higher near $T_c$ and this
may have important contribution to thermalization of the matter created
in relativistic heavy-ion collision experiments. Again there seems to be
negligible isospin dependence of $c^2_s$ in the range of temperatures
studied. From Fig.\ \ref{fg.cs} we note that in the PNJL model even with
$\mu_0$ as large as $0.8 T_c$, the $v^2_s$ never reaches a value as
large as $0.2$ near or below $T=T_c$ which was used in \cite{beda} to
describe the rapidity spectra.

\section{Discussions and Summary} \label{sc.summary}

We have extended the PNJL model of Ref.\  \cite{pnjl2} by the introduction
of isospin chemical potential. Using this we have studied the behaviour
of strongly interacting matter with two degenerate quark flavours in the
phase space of $T$, $\mu_0$ and $\mu_I$, for small values of the
chemical potentials.  We have extracted $10$ coefficients of Taylor
expansion of pressure in the two chemical potentials upto $6$-th order.
Some of these coefficients were compared with available LQCD data. The
quark number susceptibility and isospin susceptibility show
order parameter-like behaviour. A quantitative comparison shows that the
quark number susceptibility reaches about $85 \%$ of its ideal gas value
upto temperature of about $2T_c$, consistent with LQCD results. However,
the isospin susceptibility reaches its ideal gas value by this
temperature. This is in contrast to LQCD results where both the
susceptibilities are almost equal from around $1.2T_c$ onward.
Similarly, the higher order derivatives for $\mu_I$ approach the the
ideal gas behaviour much faster compared to those for $\mu_0$.  In
contrast, though both the QNS and INS in LQCD deviate from their ideal
gas values, the higher order derivatives reach their ideal gas limit
quickly.  The values of the mixed derivatives in the PNJL model shows a
behaviour somewhat in between. On the Lattice however, the mixed
derivatives are almost zero (\ie, the ideal gas value) above $T_c$.

Thus some of the coefficients in the PNJL model differ from the LQCD
data and one could hold the mean field analysis responsible for this
departure. But if this argument were true then the  higher order
derivatives obtained in the PNJL model should depart from the LQCD data
more than the lower order coefficients, which is not the case. As
against this expectation, we have found a very nice pattern in the PNJL
results which can be understood in terms of the behaviour of the
Polyakov loop. The dependence of the Polyakov loop and its conjugate on
temperature and the chemical potentials is extremely important. For
$\mu_I=0$ they have different values when $\mu_0$ is varied. This makes
all the coefficients which are derivatives of pressure with respect to
$\mu_0$ alone, to deviate from the ideal gas behaviour. For $\mu_0 =0$
however the Polyakov loop and its conjugate are equal and hence both
reach the ideal gas value above $T_c$. Thus the coefficients which are
derivatives of pressure with respect to $\mu_I$ alone, all reach their
respective ideal gas values above $T_c$. The mixed derivatives are found
to be somewhere in between. Nonetheless, we hope to look into the effects
of fluctuations in future.

In order to have a closer look at the discrepancy between the PNJL
results and LQCD data, we have also calculated the flavour diagonal and
flavour off-diagonal susceptibilities upto $6$-th order. We have found
that the $2$-nd order flavour off-diagonal susceptibility, which
indicates the correlation among ``up'' and ``down'' quarks, is significantly
away from zero even upto $T = 2 T_c$. On the other hand, LQCD results 
\cite{swagato} show that correlation among the flavours in the $2$-nd 
order off-diagonal susceptibility is largely governed by the interaction 
of the quarks with the gauge fields and is almost independent of the 
presence of the quarks loops.  This motivated us to study the
instanton induced flavour-mixing effects within the framework of the
PNJL model. Unfortunately, we found no effect of flavour-mixing on any
diagonal QNS and INS, and hence on the $2$-nd order flavour off-diagonal
susceptibility. We speculate one possibility to reconcile PNJL and 
LQCD data, that is to keep the pion condensate as a dynamical variable 
and perform the calculations. In fact there are indications \cite{kli3}
that at zero temperature and in the chiral limit, pion condensation can
be catalysed by an external chromomagnetic field. We hope to present 
these results in future.

On the other
hand, flavour-mixing effects on the mixed susceptibilities of quark and
isospin chemical potentials indicate that large flavour-mixing is
favoured by the LQCD data. This may have important consequences
\cite{buballa} on the phase diagram of the NJL model at low temperature
and large baryon chemical potential.

Apart from the possible improvements for the Polyakov loop potential,
inclusion of pionic and diquark fluctuations, etc. we also intend
to include terms in the NJL part with six point couplings to take
proper account of the quark number fluctuations in the low temperature
phase.

We have also investigated chemical potential dependence of specific heat
and speed of sound. The specific heat sort of becomes independent of
chemical potential just above $T_c$. Below $T_c$ there is some
significant effect from both the chemical potentials $\mu_0$ and
$\mu_I$. Consistent with LQCD findings \cite{allton,kogut1}, the peak in
specific heat  towards lower temperatures with increasing chemical
potentials indicating a decrease in the transition temperature. We plan
to make a more detailed investigation of the location of the phase
boundary. The speed of sound, on the other hand, increases with the
increase of either of the chemical potentials in almost the whole range
of temperatures. But this dependence become milder as one goes to higher
temperatures. Thus with a proper implementation of the PNJL equation of
state into the hydrodynamic studies of elliptic flow, one may able to
make some estimates of both the temperature and densities reached in the
heavy-ion collision experiments.

\begin{acknowledgments}
RR would like to thank S. Digal for many useful discussions and comments.
\end{acknowledgments}

\appendix 

\section{} \label{ap.model}
Here we present the complete details of the PNJL model used in our work.
First we discuss the NJL model in the complete space of temperature $T$
and the ``up" and ``down" flavour chemical potentials $\mu_u$ and $\mu_d$ 
(or equivalently the quark chemical potential $\mu_0$ and isospin
chemical potential $\mu_I$) \cite{asakawa1,buballa}. Then we extend it
to couple with the Polyakov Loop.

\subsection{The NJL model}
The NJL model Lagrangian for two flavours can be written as 
\cite{asakawa1,buballa}:
\bsubeq
\beqa
  {\mathcal L} &=& {\mathcal L}_0 + {\mathcal L}_1 + {\mathcal L}_2 ~~~,\\ 
  {\mathcal L}_0 &=& \bar{\psi}(i\dslash-m)\psi ~~~,\\
  {\mathcal L}_1 &=& G_1\l[ \l(\bar{\psi}\psi\r)^2 +
  \l(\bar{\psi}\vec{\tau}\psi\r)^2  +  \l(\bar{\psi}i\gamma_5\psi\r)^2  +
  \l(\bar{\psi}i\gamma_5\vec{\tau}\psi\r)^2 \r] ~~~,\\
  {\mathcal L}_2 &=& G_2\l[ \l(\bar{\psi}\psi\r)^2 -
  \l(\bar{\psi}\vec{\tau}\psi\r)^2  -  \l(\bar{\psi}i\gamma_5\psi\r)^2  +
  \l(\bar{\psi}i\gamma_5\vec{\tau}\psi\r)^2 \r] ~~~,
\eeqa
\label{eq.lagrangian}
\esubeq
where,
\beq
 \psi=(u,d)^T, \qquad\qquad [G_1]=[G_2]=[{\rm energy}]^{-2}, \qquad\qquad
 m=diag(m_u,m_d). 
\eeq
We shall assume flavour degeneracy $m_u=m_d=m_0$.
For $m_0 = 0$ the symmetries of the different parts of the Lagrangian
\ref{eq.lagrangian} are:
\bsubeq
\beqa
  {\mathcal L}_0 &:& SU_V(2)\times SU_A(2)\times U_V(1)\times U_A(1) \\
  {\mathcal L}_1 &:& SU_V(2)\times SU_A(2)\times U_V(1)\times U_A(1) \\
  {\mathcal L}_2 &:& SU_V(2)\times SU_A(2)\times U_V(1)
\eeqa
\esubeq
${\mathcal L}_2$ has the structure of a 't-Hooft determinant, ${\rm
det}\l[{\bar q}(1+\gamma_5)q\r]+{\rm det}\l[{\bar q}(1-\gamma_5)q\r]$
\cite{njls}, and breaks $U_A(1)$ axial symmetry. This interaction
can be interpreted as induced by instantons and reflects the
$U_A(1)$-anomaly of QCD.

We are interested in the properties of this Lagrangian at nonzero
temperatures $T$ and chemical potentials $\mu_u$ and $\mu_d$. 
Equivalently, one can also use the quark number chemical potential
$\mu_0 = (\mu_u+\mu_d)/2$ and the isospin chemical potential
$\mu_I = (\mu_u-\mu_d)/2$. In the mean field approximation we 
consider the two quark condensates $\sigma_u = \la{\bar u}u\ra$ 
and $\sigma_d=\la{\bar d}d\ra$. The pion condensate $\vec{\pi}$ is 
assumed to be zero (which is true in the NJL model for 
$\mu_I < m_{\pi}/2$). Then the thermodynamic potential is obtained as,

\bsubeq
\beqa
 \Omega(T,\mu_u,\mu_d) &=& \sum_{f=u,d}\Omega_0(T,\mu_f;m_f) +
 2G_1\l(\sigma_u^2+\sigma_d^2\r) + 4G_2\sigma_u\sigma_d ~~~, \\ 
 \Omega_0(T,\mu_f;m_f) &=& -2N_c\int\frac{d^3p}{\l(2\pi\r)^3} 
    E_f\theta(\Lambda^2-\vec{p}~^2) -
    2N_cT\int\frac{d^3p}{\l(2\pi\r)^3}\ln\l[1+e^{-(E_f-\mu_f)/T}\r]
    \nonumber \\
    &&-2N_cT\int\frac{d^3p}{\l(2\pi\r)^3}\ln\l[1+e^{-(E_f+\mu_f)/T}\r] ~~~.
\eeqa
\label{eq.potential} 
\esubeq

where, the energy $E_f$ and constituent quark mass $m_f$ is given
by,

\beqa
E_f &=& \sqrt{m_f^2+p^2} ~~, \\
m_f &=& m_0 - 4G_1\sigma_f - 4G_2\sigma_{f'}, \qquad\qquad f\ne
  f'\in\{u,d\} ~~.
\label{eq.mass}  
\eeqa

Finding the stationary points of the thermodynamic potential with
respect to $\sigma_u$ and $\sigma_d$, \ie, solving the coupled equations
$\partial\Omega/\partial\sigma_u=0$ and
$\partial\Omega/\partial\sigma_d=0$, one gets the gap equations,
\bsubeq
\beqa
  \sigma_f = -2N_c\int\frac{d^3p}{\l(2\pi\r)^3}\frac{m_f}{E_f}
    \l[ \theta(\Lambda^2-p^2)- n(E_f) - {\bar n}(E_f) \r],
    \qquad f = {u,d} \\
    n(E_f)=\frac{1}{1+\exp(E_f-\mu_f)}, \qquad{\rm and}\qquad 
  {\bar n}(E_f)=\frac{1}{1+\exp(E_f+\mu_f)} .  
\eeqa
\label{eq.condensate}
\esubeq
The constituent mass $m_f$ for one flavour depends in general on both
the condensates [see Eqn.\ (\ref{eq.mass})] and therefore the two flavours
are coupled. Chiral symmetry ($SU_A(2)$) is broken spontaneously for
$\sigma_f \ne 0$.\\
Let us now make the parameterization 
\beq
  G_1=(1-\alpha)G_0, \qquad{\rm and}\qquad G_2=\alpha G_0
\eeq
with a fixed value of $G_0$. Tuning the value of $\alpha$ one can
control the flavour mixing in the Lagrangian. We consider some of the
cases below.
\begin{enumerate}
\item \underline{$\alpha=0$} : This implies $G_2=0$ \ie the $U_A(1)$
symmetry breaking term ${\mathcal L}_2$ drops out and hence 
${\mathcal L}$ has no flavour mixing. 

\item \underline{$\alpha=1$} : Here $G_1=0$, and thus ${\mathcal L}_2$
completely dominates the coupling. The flavour mixing in 
${\mathcal L}$ is thus ``maximal''. 

\item \underline{$\alpha=1/2$} : In this case we have $G_1=G_2=G_0/2$. 
So the Lagrangian ${\mathcal L}={\mathcal L}_0 + 
G_0\l[ \l(\bar{\psi}\psi\r)^2 + 
\l(\bar{\psi}i\gamma_5\vec{\tau}\psi\r)^2 \r]$, is the standard NJL model
\cite{njl1}. Here also the $U_A(1)$ symmetry is broken which is commensurate
with the fact that in nature the $\eta$ particle is much heavier than
the $\pi$'s

\end{enumerate}
\subsection{Extension to PNJL}
Our aim is to extend the PNJL model introduced in Ref.\ .\cite{pnjl1,pnjl2}
to include isospin chemical potential. To achieve this we now include the
Polyakov loop and its effective potential to the NJL model described
above. The Lagrangian becomes,
\beqa
  {\mathcal L}_{PNJL} &=& {\mathcal L}_0 + {\mathcal L}_1 + {\mathcal
  L}_2 - {\mathcal U}\l( \Phi[A],\bar{\Phi}[A],T \r) 
\eeqa

\noindent
The only part of the NJL sector that is modified is ${\mathcal L}_0$ 
which now becomes,
\beqa
  {\mathcal L}_0 = \bar{\psi}\l( i\Dslash - m \r)\psi ~~~,
\eeqa
\noindent
where 
\beqa
  D^\mu=\partial^\mu-iA^\mu, \qquad A^\mu=\delta_{\mu0}A^0, \qquad
  A^\mu(x)=g{\mathcal A}^\mu_a(x)\lambda_a/2 .
\eeqa

\noindent
${\mathcal A}^\mu_a(x)$ are $SU(3)$ gauge fields and $\lambda_a$ are
Gell-Mann matrices. 

${\mathcal U}\l( \Phi[A],\bar{\Phi}[A],T \r)$ is the effective potential
expressed in terms of the traced (over color) Polyakov loop (with
periodic boundary conditions) and its charge conjugate---
\beq
  \Phi = \frac{{\rm Tr}_c L }{N_c}, \qquad
  \bar{\Phi} = \frac{{\rm Tr}_c L^\dagger }{N_c}, \qquad
  L(\vec{x}) = {\mathcal P}\exp\l[i\int_0^\beta d\tau\
  A_4(\vec{x},\tau)\r],
  \qquad \beta=\frac{1}{T}, \qquad A_4=iA_0 .
\eeq

We shall be working in the mean field limit. For simplicity of notation
we shall use $\Phi$ and $\bar{\Phi}$ as their respective mean fields.
$\Phi$ is the order parameter for deconfinement transition. In the 
absence of quarks $\Phi=\bar{\Phi}$ and deconfinement is associated
with the spontaneous breaking of the $Z(3)$ symmetry. Conforming to
this symmetry and parameterizing the LQCD Monte Carlo data one
can write down an effective potential for $\Phi$ and $\bar{\Phi}$.
Following Ref.\ \cite{pnjl2}, we write
\bsubeq
\beqa
  \frac{{\mathcal U}(\Phi,\bar{\Phi},T)}{T^4} &=& 
    -\frac{b_2(T)}{2}\Phi\bar{\Phi} -
    \frac{b_3}{6}\l(\Phi^3+{\bar{\Phi}}^3\r) +
    \frac{b_4}{4}\l(\bar{\Phi}\Phi\r)^2, \\
  b_2(T) &=& a_0 + a_1\l(\frac{T_0}{T}\r) + a_2\l(\frac{T_0}{T}\r)^2 +
    a_3\l(\frac{T_0}{T}\r)^3 .
\eeqa
\esubeq
At low temperature ${\mathcal U}$ has a single minimum at $\Phi=0$,
while at high temperatures it develops a second one which turns into the
absolute minimum above a critical temperature $T_0$. $\Phi$ and
$\bar{\Phi}$ will be treated as independent classical fields.
The mean field analysis of the NJL part of the model proceeds in exactly
the same way as in the previous case. Using $\sigma_u$ and $\sigma_d$ as
the independent quark condensates (and neglecting $\vec{\pi}$) one gets
the expression for the thermodynamic potential,

\bsubeq
\beqa
 \Omega(T,\mu_u,\mu_d) &=&  {\mathcal U}\l(\Phi,\bar{\Phi},T\r) + 
  \sum_{f=u,d}\Omega_0(T,\mu_f;M_f) +
  2G_1\l(\sigma_u^2+\sigma_d^2\r) + 4G_2\sigma_u\sigma_d ~~~, \\ 
 \Omega_0(T,\mu_f;M_f) &=& -2N_c\int\frac{d^3p}{\l(2\pi\r)^3} 
    E_f\theta(\Lambda^2-\vec{p}~^2) -
    2T\int\frac{d^3p}{\l(2\pi\r)^3}{\rm
    Tr}_c\ln\l[1+Le^{-(E_f-\mu_f)/T}\r]
    \nonumber \\
    &&-2T\int\frac{d^3p}{\l(2\pi\r)^3}{\rm
    Tr}_c\ln\l[1+L^\dagger e^{-(E_f+\mu_f)/T}\r] ~~~.
\eeqa
\label{eq.potential1} 
\esubeq

\noindent
where $ E_{u,d} = \sqrt{m_{u,d}^2+p^2}$ and
 $m_{u,d} = m_0 - 4G_1\sigma_{u,d} - 4G_2\sigma_{d,u}$.

Note that with $\alpha=0.5$ and $\mu_u=\mu_d=\mu$ and if for the coupling 
$G$ and condensate $\sigma$ of Ref.\ \cite{pnjl2} one uses $G=2G_0$ and 
$\sigma=\sigma_u+\sigma_d$, then the thermodynamic potentials here and
in Ref.\ \cite{pnjl2} are exactly equal.

Using the identity ${\rm Tr}\ln X = \ln{\rm det}X$ one can write for a
given flavour $f$,

\beqa
  && \ln{\rm det}\l[1+Le^{-(E_f-\mu_f)/T}\r] + \ln{\rm det}\l[1+L^\dagger
  e^{-(E_f+\mu_f)/T}\r] \nonumber \\ && = \ln\l[ 1 + 
  3\l(\Phi+\bar{\Phi}e^{-(E_f-\mu_f)/T}\r)e^{-(E_f-\mu_f)/T} +
  e^{-3(E_f-\mu_f)/T} \r] \nonumber \\ && + \ln\l[ 1 +
  3\l(\bar{\Phi}+\Phi e^{-(E_f+\mu_f)/T}\r)e^{-(E_f+\mu_f)/T} +
  e^{-3(E_f+\mu_f)/T} \r] .
\eeqa

This gives us the final form of the thermodynamic potential as,
\bsubeq
\beqa
 \Omega(T,\mu_u,\mu_d) &=&  {\mathcal U}\l(\Phi,\bar{\Phi},T\r) + 
  \sum_{f=u,d}\Omega_0(T,\mu_f;M_f) +
  2G_1\l(\sigma_u^2+\sigma_d^2\r) + 4G_2\sigma_u\sigma_d, \\ 
 \Omega_0(T,\mu_f;M_f) &=& -2N_c\int\frac{d^3p}{\l(2\pi\r)^3} 
    E_f\theta(\Lambda^2-\vec{p}~^2) \nonumber \\ 
    && -2T\int\frac{d^3p}{\l(2\pi\r)^3} \ln\l[ 1 +
    3\l(\Phi+\bar{\Phi}e^{-(E_f-\mu_f)/T}\r)e^{-(E_f-\mu_f)/T} +
    e^{-3(E_f-\mu_f)/T} \r] \nonumber \\
    && -2T\int\frac{d^3p}{\l(2\pi\r)^3} \ln\l[ 1 +
    3\l(\bar{\Phi}+\Phi e^{-(E_f+\mu_f)/T}\r)e^{-(E_f+\mu_f)/T} +
    e^{-3(E_f+\mu_f)/T} \r] ~~~.
\eeqa
\label{eq.potential2} 
\esubeq
From this thermodynamic potential the
equations of motion for the mean fields $\sigma_u$, $\sigma_d$, $\Phi$
and $\bar{\Phi}$ are derived through,
\beq
  \frac{\partial\Omega}{\partial\sigma_u}=0, \qquad
  \frac{\partial\Omega}{\partial\sigma_d}=0, \qquad
  \frac{\partial\Omega}{\partial\Phi}=0, \qquad
  \frac{\partial\Omega}{\partial\bar{\Phi}}=0 .
\eeq
This coupled equations are then solved for the fields as functions 
of $T$, $\mu_u$ and $\mu_d$. They give, 
\bsubeq
\beqa
  \sigma_f &=& -6\int\frac{d^3p}{(2\pi)^3}\frac{m_f}{E_f}\l[
  \theta(\Lambda^2-p^2) - {\mathcal N}(E_f){\mathcal M}(E_f) -
  \bar{\mathcal N}(E_f)\bar{\mathcal M}(E_f) \r]; 
  \qquad f=u,d \\
  \frac{\partial{\mathcal U}}{\partial\Phi} &=& 6T\sum_{f=u,d}
  \int\frac{d^3p}{(2\pi)^3}\l[ {\mathcal N}(E_f)e^{-(E_f-\mu_f)/T} +
  \bar{\mathcal N}(E_f)e^{-2(E_f+\mu_f)/T} \r], \\  
  \frac{\partial{\mathcal U}}{\partial\bar{\Phi}} &=& 6T\sum_{f=u,d}
  \int\frac{d^3p}{(2\pi)^3}\l[ {\mathcal N}(E_f)e^{-2(E_f-\mu_f)/T} +
  \bar{\mathcal N}(E_f)e^{-(E_f+\mu_f)/T} \r], \\  
  {\mathcal N}(E_f)&=& \l[1 +
    3\l(\Phi+\bar{\Phi}e^{-(E_f-\mu_f)/T}\r)e^{-(E_f-\mu_f)/T} +
    e^{-3(E_f-\mu_f)/T} \r]^{-1}, \\  
  \bar{\mathcal N}(E_f)&=& \l[1 +
    3\l(\bar{\Phi}+\Phi e^{-(E_f+\mu_f)/T}\r)e^{-(E_f+\mu_f)/T} +
    e^{-3(E_f+\mu_f)/T} \r]^{-1}, \\  
  {\mathcal M}(E_f) &=&
  \l(\Phi+2\bar{\Phi}e^{-(E_f-\mu_f)/T}\r)e^{-(E_f-\mu_f)/T} +
  e^{-3(E_f-\mu_f)/T}, \\
  \bar{\mathcal M}(E_f) &=& \l(\bar{\Phi}+2\Phi
  e^{-(E_f+\mu_f)/T}\r)e^{-(E_f+\mu_f)/T} + e^{-3(E_f+\mu_f)/T} .
\eeqa \label{eq.sig-u-d}\
\esubeq
Finally we note that the values of the parameters used are exactly
the same as used in Ref.\ \cite{pnjl3}.

%%%%%%%%%%%%%%%%%%%%%%%%%%%%%%%%%%%%%%%%%%%%%%%%%%%%%%%%%%%%%%%%%%%%%%%

\end{document}